\documentclass[a4paper,11pt]{article}
\pdfoutput=1 

\usepackage{jcappub} 
\usepackage[T1]{fontenc} 

\usepackage{graphicx}
\usepackage{graphicx,color}
\usepackage{amssymb}
\usepackage{amsmath}
\usepackage{bm}
\usepackage{hyperref}
\usepackage{multirow}
\usepackage{longtable}

\usepackage{float}
\usepackage[normalem]{ulem}
\restylefloat{table}

\usepackage[dvipsnames]{xcolor}

\begin{document}

\title{Testing the growth rate in homogeneous and inhomogeneous interacting vacuum models}

\author[a]{H. A. Borges,}
\author[a]{C. Pigozzo,}
\author[a]{P. Hepp,}
\author[a,b]{L. O. Baraúna,}
\author[c,d]{M. Benetti}

\affiliation[a]{Universidade Federal da Bahia, Salvador, BA, 40210-340, Brazil} 

\affiliation[c]{Instituto Nacional de Pesquisas Espaciais, 
São José dos Campos, SP 12227-900, Brazil}

\affiliation[c]{Scuola Superiore Meridionale,  Largo San Marcellino 10, 80138 Napoli, Italy}
\affiliation[d]{Istituto Nazionale di Fisica Nucleare (INFN), Sezione di Napoli, Via Cintia, 80126 Napoli, Italy}

\emailAdd{humberto@ufba.br}
\emailAdd{cpigozzo@ufba.br}
\emailAdd{patiheppx@gmail.com}
\emailAdd{luanorion1@gmail.br}
\emailAdd{micol.benetti@unina.it}

\abstract{

In this work we consider a class of interacting vacuum corresponding to a generalised Chaplygin gas (gCg) cosmology. 
In particular we analyse two different scenarios at perturbation level for the same background interaction characterised by the parameter $\alpha$: (i) matter that follows geodesics, corresponding to homogeneous vacuum, and (ii) a covariant ansatz for vacuum density perturbations. 
In the latter case, we show that the vacuum perturbations are very tiny as compared to matter perturbations on sub-horizon scales. In spite of that, depending on the value of the Chaplygin gas parameter $\alpha$, vacuum perturbations suppress or enhance the matter growth rate as compared to the case (i).
We use Cosmic Microwave Background (CMB), type Ia supernovae (SNe) and Redshift Space Distortion (RSD) measurements to test the observational viability of the model. We found that the mean value of our joint analysis clearly favours a positive interaction, i.e., an energy flux from dark matter to dark energy, with $\alpha \approx 0.143$ in both cases, while the cosmological standard model, recovered for $\alpha$=0, is ruled out by 3$\sigma$ confidence level. Noteworthy, the positive value of interaction can alleviate both the $H_0$ and $S_8$ tension for the dataset considered here.
}

\maketitle

\section{Introduction}\label{intro}

The accelerated expansion of the universe detected by the precision measurements of type Ia supernovae \cite{Astier,riess,perl}, anisotropies in the cosmic microwave background radiation \cite{spergel, g, ade, ade1} and baryon acoustic oscillations \cite{telubac, ata} indicates that about $96\%$ of the total energy of our universe is in the form of unknown dark fluid components, namely, dark energy \cite{tegmark} and dark matter, and the remaining components are in the form of baryonic matter and radiation. The clustering dark matter with zero pressure (cold dark matter) is concentrated in the local structures and plays crucial role for forming galaxies and clusters of galaxies while dark energy is featured by a negative pressure that drives the recent accelerated expansion. The simplest model to describe dark energy is a constant vacuum energy density $\rho_V$ characterised by the equation of state parameter $w=-1$, equivalent to a cosmological constant in Einstein gravity \cite{pebles,Padmanaban,weinberg02}. The cosmological model that incorporates a constant vacuum energy plus cold dark matter is known as $\Lambda$CDM. Despite its relative success when tested against the most precise observations, one of its problems is the large discrepancy between the current vacuum energy density, $\rho_V\sim 10^{-29} g/cm^3$, and the theoretical value predicted by quantum field theories \cite{weinberg02}. 

Among of many alternatives to circumvent the problem of the cosmological constant we can consider the generalised Chaplygin gas (gCg), a unified dark sector whose equation of state is given by \cite{kam, Fabris, Bento, Sandvik:2002jz, bento} 
\begin{equation}\label{g}
p=-\frac{A}{\rho^{\alpha}},
\end{equation}
where A is a constant, $\rho$ is the energy density and $\alpha$ is a free parameter.  This gCg interpolates between a cold dark matter dominating at early times and a dark energy component dominating at late times. However, due to a non-zero adiabatic sound speed, the perturbations exhibit strong instabilities and oscillations affecting radically the matter power spectrum, unless the parameter $\alpha$ does not differ too much from zero. An alternative to avoid this problem is to include a non-adiabatic pressure contribution in order to make the effective sound speed vanish \cite{amendola}. Another possibility is to split the gCg fluid into two interacting components, a pressuless cold dark matter with energy density $\rho_m$ and a vacuum term with equation of state $p_V = - \rho_V$ \cite{Zi, Wa, Ba}. In this way, the sound speed comes only from the vacuum energy perturbations, that need to be zero or negligible as compared to matter perturbations.

The above condition is reached if we assume that dark matter follows geodesics, with energy transfer proportional to its $4$-velocity, which implies that energy and momentum exchange between vacuum and CDM is absent at perturbative level \cite{DW, DW1}. However, due to the dynamical nature of vacuum, it is important to consider the possibility of inhomogeneities in its energy density and verify explicitly if their fluctuations are in fact negligible at sub-horizon scales. This was done in a model equivalent to the generalised Chaplygin gas with $\alpha=$-1/2 \cite{W}, but here we extend the analysis for any value of $\alpha$.

Since we are interested to explore the impact of vacuum energy fluctuations on the evolution of the dark matter growth rate in a decomposed gCg model with interacting dark matter and vacuum energy, for the sake of simplicity, we neglect the contributions of the baryonic component at first moment. Under the assumption of a covariant ansatz for the vacuum energy density, we are able to compute the energy-momentum transfer up to first order using a gauge invariant perturbative approach. A scale-dependent second order differential equation for the  matter density contrast is obtained, which allows us to follow the evolution of the linear dark matter growth rate, defined by
\begin{equation}\label{f}
f=\frac{\dot\delta_m}{H\delta_m}.
\end{equation}
We show explicitly that the vacuum perturbations are smaller than the matter fluctuations by several orders of magnitude on sub-horizon scales. However, due to the corrections in the second order differential equation for the matter density contrast we show that the growth rate evolution at late time is not the same evolution as found in the geodesic scenario.

Since the observed matter power spectrum $P(k)$ and the redshift space distortion (RSD) is a measure of the galaxy distribution, the conserved baryons will be included in the perturbative dynamics.

In light of observations, previous analysis considering Cosmic Microwave Background measurements (CMB) from Planck 2015 with the SNeIa data from the JLA sample are consistent with values very close to the $\Lambda$CDM(i.e. zero $\alpha$), with $|\alpha|\leq 0.05$ \cite{W. Luciano}. On the other hand, considering the lensing amplitude, $\mathcal{A}_l$, and the effective number of relativistic species, $N_{eff}$, as a free parameters, the data favour negative values of $\alpha$, reducing the discrepancy between the local and CMB $H_0$ measurements \cite{Benetti:2019lxu}. The update results using CMB measurements from Planck 2018, SNeIa data and SH0ES prior, assuming a spatially flat universe, favours a positive interaction with $\alpha\approx$ 0.2, ruled out $\Lambda$CDM model by more than 3$\sigma$ confidence level. The influences of non-zero spatial curvature and SH0ES prior on $H_0$ in the joint analyses are also considered \cite{Benetti:2021lxu}.

Motivated by such a previous results, in this work we take a step forward and we generalize the analysis done in \cite{W} not assuming $\alpha$ at a specific value but leaving it free to vary, including the dynamics of the baryonic component to be able to compute RSD and the observed matter power spectrum. We also perform an updated analysis of \cite{W. Luciano} and \cite{Benetti:2021lxu} to take into account RSD data. 

This paper is organised as follows. In section II we present the background evolution of the decomposed gCg model. In section III we perform a perturbative analysis of the evolution of the dark matter growth rate sourced by vacuum energy perturbations. A comparison is then made with the case of geodesic scenario. In section IV we present the observational tests and the results. Finally, we conclude the study in Sec. V. 

\section{Background model}

In this section we present the background evolution of the decomposed gCg model as the interacting model. For this aim, let us consider a flat FLRW cosmology in which the cosmic interacting fluid is described by the balance equations 

\begin{eqnarray}\label{continuity3}
\dot\rho_m + 3H\rho_m =Q, \\
\dot\rho_{V}=-Q,
\end{eqnarray}
where $\rho_m$ and $\rho_V$ are respectively the densities of pressureless matter and vacuum. The energy transfer between the components is given by
\begin{equation}\label{Gamma}
Q = 6\alpha H_0(1-\Omega_{m 0})\bigg(\frac{H}{H_0}\bigg)^{-(2\alpha+1)}\dot{H}.
\end{equation}

This model corresponds to the decomposed gCg \cite{kam, Fabris, Bento, Sandvik:2002jz, bento, bilic, JSA} whose pressure is given by ($\ref{g}$). The sign of $Q$ depends on the sign of the gCg parameter $\alpha$, since $\dot H<0$. If $Q>0$ the vacuum energy decays into matter, and in the opposite case matter decays into vacuum energy. Therefore, we see that, for $\alpha < 0$, the vacuum energy density decays along the expansion, while matter is created in the process, whereas for $\alpha > 0$ matter is annihilated. 

On the other hand, for $\alpha = 0$ we re-obtain the standard model with a cosmological constant and conserved matter.

The solutions for matter and vacuum densities are, respectively
\begin{equation} \label{Matter}
\rho_{m}=3H^2-3H_{0}^{2(1+\alpha)}(1-\Omega_{m0})H^{-2\alpha},
\end{equation}
\begin{equation}\label{pl}
\rho_{V}=\rho_{V 0}\bigg(\frac{H}{H_0}\bigg)^{-2\alpha},
\end{equation}
where a subindex 0 indicates the present value of the corresponding quantities and $\Omega_{m 0}=\rho_{m0}/3H_{0}^2$ is the present matter density parameter.

The Hubble expansion rate is governed by the Friedmann equation \cite{ca}
\begin{equation}
H = H_0\sqrt{\bigg[1-\Omega_{m0}+\frac{\Omega_{m0}}{a^{3(1+\alpha)}}\bigg]^{\frac{1}{1+\alpha}}+\frac{\Omega_{r0}}{a^4}}.
\end{equation}

\subsection{The covariant energy transfer}

We can explicitly write the energy transfer term ({\ref{Gamma}}) in a covariant manner for a perfect fluid described by two interacting components. For this purpose, let us assume a covariant form for the vacuum energy density ($\ref{pl}$) \cite{CPS, W. Luciano}, 
\begin{equation}\label{pls}
\rho_{V}=\rho_{V 0}\bigg(\frac{\Theta}{3H_0}\bigg)^{-2\alpha},
\end{equation}
where we use the scalar expansion $\Theta=u^{\mu}_{;\mu}$ with $u^{\mu}$ being the four velocity of the fluid. In the background universe the scalar expansion is $\Theta = 3H$.

The energy-momentum conservation equations for each component are given by
\begin{equation}\label{asd}
{T^{\nu\mu}_m}_{;\mu}=Q^{\nu},
\end{equation}
\begin{equation}\label{asdw}
{T^{\nu\mu}_V}_{;\mu}=-Q^{\nu},
\end{equation}
where
\begin{equation}
 T^{\mu\nu}_A=\rho_Au^{\mu}u^{\nu}+p_Ah^{\mu\nu}
 \end{equation}
is the energy-momentum tensor of each component, $h^{\mu\nu}=g^{\mu\nu}+u^{\mu}u^{\nu}$ is the projector tensor and $Q^{\mu}$ is the energy-momentum transfer between dark matter and vacuum. The latter can be decomposed parallel and perpendicular to the four velocity $u^{\mu}$ as
\begin{equation}\label{dec}
Q^{\mu}=u^{\mu}Q+\bar{Q}^{\mu},
\end{equation}
with $Q=-u_{\mu}Q^{\mu}$, $\bar{Q}^{\mu}=h^{\mu}_{\nu}Q^{\nu}$, $u_{\mu}\bar{Q}^{\mu}=0$ and $u_{\mu}u^{\mu}=-1$.

Projecting equations $(\ref{asd})$ and $(\ref{asdw})$ parallel and perpendicular to $u^{\mu}$, we find the energy conservation equations
\begin{equation}\label{lopq}
\rho_{m,\mu}u^{\mu}+\Theta\rho_m=-u_{\mu}Q^{\mu},
\end{equation}
\begin{equation}\label{lapis}
\rho_{V,\mu}u^{\mu}=u_{\mu}Q^{\mu},
\end{equation}
and the momentum conservation equations
\begin{equation}\label{hum}
\rho_m{u^{\mu}}_{;\nu}u^{\nu}=\bar{Q}^{\mu},
\end{equation}
\begin{equation}\label{huma}
\rho_{V,\nu}h^{\nu\mu}=-\bar{Q}^{\mu}.
\end{equation}
Using the ansatz $(\ref{pls})$ in the last equation, we find explicitly the covariant energy transfer source term
\begin{equation}\label{werb}
Q=\frac{2}{3}\alpha(1-\Omega_{m 0})(3H_0)^{2(\alpha+1)}\Theta^{-(2\alpha+1)}{\Theta}_{,\mu}u^{\mu}.
\end{equation}
We complete our system of equations with the Raychaudhuri equation
\begin{equation}\label{sabe}
\Theta_{,\mu}u^{\mu}=-\frac{1}{3}\Theta^2+(u^{\mu}_{;\nu}u^{\nu})_{;{\mu}}-\frac{1}{2}(\rho_m-2\rho_V),
\end{equation}
where we have neglected the shear and vorticity contributions.
In the comoving frame, where the components of the four-velocity are $u_{0}=-1$, $u^{0}=1$ and $u_{i}=0=u^{i}$, one has $\bar{Q}^{\mu}=0$, which shows that there is no momentum transfer in the homogeneous and isotropic background and the expression ({\ref{Gamma}}) is recovered.

\section{The perturbed equations}

Now let us focus our attention to linear perturbations around a spatially flat FLRW universe. Let us start with the most general line element for scalar perturbations \cite{JB},
\begin{equation}
ds^2=-(1+2\phi)dt^2+2a^2B_{,i}dtdx^i+a^2[(1-2\psi)\delta_{ij}dx^idx^j+2E_{,ij}dx^idx^j].
\end{equation}
We follow the approach presented in the reference \cite{W} whose analysis is gauge-invariant, here we generalise the results for any value of $\alpha$. At the moment we neglect the perturbation dynamics for the baryon component. 
The fluid velocity potencial $v$ can be defined by perturbing the four velocity $u_{\mu}=g_{\mu\nu}u^{\nu}$, which results in
\begin{equation}
\delta u_j=a^2\delta u^j+a^2B_j=v_{,j},
\end{equation}
assuming that $v$ is irrotational. We postulate that it coincides with the matter velocity potential $v_m$, since we cannot properly define the four velocity for the vacuum component.
The time component of the perturbed four-velocity is related to the perturbed metric through
\begin{equation}\label{estes}
\delta u_0=\delta u^0=-\phi.
\end{equation}

The next step is to obtain the conservation equations for each interacting component, namely matter and vacuum energy.
In order to provide a set of basic equations to calculate the matter density perturbation $\delta_m=\delta\rho_m/\rho_m$, we start by considering the equations for the vacuum.  The perturbation of the momentum equation $(\ref{huma})$ yields, in the comoving gauge, the result
\begin{equation}\label{momentum}
\partial^i\delta p_V^c = -\partial^i\delta \rho_V^c = -\delta\bar Q^{ic}.
\end{equation}

So, a non-zero momentum transfer $\delta\bar Q^i$ is related to the presence of spatial variations of vacuum perturbations. Here the gauge invariant scalar quantities $\delta\mathcal A^c=\delta\mathcal A+\dot{\mathcal A}v$ that characterise perturbations on comoving hypersurfaces were introduced.

The perturbation of equation $(\ref{lapis})$ allows us to compute the energy transfer between the components, 
\begin{equation}\label{vai}
\delta Q^c= \dot\rho_V(\dot v+\phi) -\delta\dot\rho^c_V.
\end{equation}

For the matter component, the energy balance $(\ref{lopq})$ and the momentum balance $(\ref{hum})$  can be written, up to first order, respectively as
\begin{equation}\label{mn}
\dot\delta_m^c+\frac{Q}{\rho_m}\delta_m^c+\delta\Theta^c=\frac{\delta Q^c}{\rho_m}+\bigg(\frac{Q}{\rho_m}-3H\bigg)(\dot v+\phi),
\end{equation}
\begin{equation}\label{sei}
(\dot v+\phi)_{, j}=\frac{\delta\bar Q_{j}^c}{\rho_m}.
\end{equation}
The latter shows that, if the momentum transfer $\delta\bar Q_j^c$ is non zero, the matter particles are forced to deviate from their geodesic motions. This means that the evolution of the matter perturbation $\delta_m^c$ should be affected by the background evolution and the source terms owing to vacuum inhomogeneities.

To complete our system of equations, the Raychaudhuri equation for the expansion is obtained from the perturbation of $(\ref{sabe})$,
\begin{equation}\label{eq2}
\delta(\Theta_{,\mu}u^{\mu})+\frac{2}{3}\Theta\delta\Theta-\frac{\nabla^2}{a^2}(\dot v+\phi)+\frac{1}{2}(\delta\rho_m-2\delta\rho_V)=0,
\end{equation}
or in terms of gauge invariant quantities
\begin{equation}\label{opai}
\delta\dot\Theta^c+\frac{2}{3}\Theta\delta\Theta^c+\frac{1}{2}\rho_m\delta_m^c=\delta\rho_V^c+\bigg(\frac{\nabla^2}{a^2}+\dot\Theta\bigg)(\dot v+\phi).
\end{equation}
For investigating the possibility of non-zero vacuum perturbations $\delta\rho^c_{V}$ and how they affect structure formation, we need now to specify a precise form for the energy and momentum transfer.

\subsection{Homogeneous vacuum density - geodesic}

Firstly, we assume that the energy transfer between the components follows the matter velocity, $Q^{\mu}=Qu^{\mu}$. In this case the momentum transfer $\bar Q^{\mu}$ is zero at the background and perturbative levels, which implies that the dynamical vacuum is homogeneous, with $\delta\rho^c_V=0$ according to $(\ref{momentum})$. Consequently, matter particles follow geodesics in a comoving frame. Furthermore, there is no energy transfer at first order, as we can see from $(\ref{vai})$ and $(\ref{sei})$. We refer to this homogeneous interaction as geodesic scenario \cite{DW}. So, the basic equations that describe the dynamics of the matter perturbations and scalar expansion are given by $(\ref{mn})$ and $(\ref{opai})$ in the absence of source terms,
\begin{equation}\label{eq1}
\dot\delta_m^c+\frac{Q}{\rho_m}\delta_m^c+\delta\Theta^c=0,
\end{equation}
\begin{equation}\label{opai1}
\delta\dot\Theta^c+\frac{2}{3}\Theta\delta\Theta^c+\frac{1}{2}\rho_m\delta_m^c=0.
\end{equation}
These equations are the same obtained in the synchronous comoving gauge \cite{HW}, and a simpler second order differential equation for the density contrast can be found. To do that, we 
differentiate the continuity equation $(\ref{eq1})$ with respect to time and eliminate $\delta\Theta^c$ and $\delta\dot\Theta^c$ by using $(\ref{eq1})$ and $(\ref{opai1})$, to obtain the following equation 
\begin{equation}\label{sin}
\ddot\delta_m^c+\bigg[\frac{Q}{\rho_m}+2H\bigg]\dot\delta_m^c+\bigg[\frac{d}{dt}\bigg(\frac{Q}{\rho_m}\bigg)+2H\frac{Q}{\rho_m}-\frac{1}{2}\rho_m\bigg]\delta_m^c=0.
\end{equation}

\subsection{Inhomogeneous vacuum density - non geodesic}

An alternative choice is to explicitly consider inhomogeneities in the vacuum, since neglecting them may lead to false interpretations of the observations \cite{Chan}. The natural manner for calculating the energy-momentum transfer at the perturbative level is to assume the covariant ansatz $({\ref{pls}})$ for the vacuum energy density, such that the perturbation of this quantity up to first order is related to the scalar expansion through the expression
\begin{equation}\label{sei23}
\delta\rho_V^c=\frac{2Q}{3\rho_m}\delta\Theta^c.
\end{equation}

The relation above can be used into $(\ref{momentum})$ to obtain the right-hand side of the momentum equation $(\ref{sei})$, given by
\begin{equation}\label{paises4}
\delta\bar Q_j={\delta\rho_V^c}_{,j}.
\end{equation}

Perturbing $(\ref{werb})$ and using the Raychaudhuri equation $(\ref{opai})$ and the relations $(\ref{paises4})$ and $(\ref{sei23})$, it is possible to write the energy transfer function in the Fourier space as
\begin{equation}\label{open45}
\frac{\delta Q^c}{\rho_m}=\frac{Q}{3\rho_m}\delta_m^c+\bigg[2H-\frac{2Q}{3\rho_m}+\frac{2QH^2}{3\rho_m^2}\bigg(\frac{k}{aH}\bigg)^2-\frac{(2\alpha+1)\rho_m}{2H}\bigg]\frac{\delta\rho_{V}^c}{\rho_m},
\end{equation}
where $k$ is the comoving wave number. The scale dependence that appears in the third term into the brackets is due to the momentum transfer between the dark components, owing to the presence of vacuum perturbations. We refer to this scenario as non geodesic. The amplitude of these perturbations compared to the matter perturbations can be evaluated by using $(\ref{mn})$ together with $(\ref{sei})$, $(\ref{paises4})$ and $(\ref{open45})$, leading to
\begin{equation}\label{fui}
\frac{\delta\rho_{V}^c}{\delta\rho_m^c}=-\frac{2Q}{3\rho_m^2 K}\bigg[Hf+\frac{2Q}{3\rho_m}\bigg].
\end{equation} 
Here we have defined the scale dependent function
\begin{equation}\label{poise}
K(a,k)=1-\frac{2Q}{3\rho_m^2}\bigg[A-H-\frac{(2\alpha+1)\rho_m}{2H}\bigg],
\end{equation}
where
\begin{equation}
A(a,k)=\frac{Q}{3\rho_m}+\frac{2QH^2}{3\rho_m ^2}\bigg(\frac{k}{aH}\bigg)^2 .
\end{equation}

The Raychaudhuri equation $(\ref{opai})$ can be written as
\begin{equation}\label{opai3}
{\delta\dot\Theta}^c=-\frac{1}{2}\rho_m\delta_m-2H\delta\Theta^c-\bigg[\frac{H^2}{\rho_m}\bigg(\frac{k}{aH}\bigg)^2+\frac{1}{2}\bigg]\delta\rho_{V}^c.
\end{equation}
Now we can differentiate $(\ref{fui})$, eliminate $\delta \dot{\Theta}^c$ through the perturbed Raychaudhuri equation $(\ref{opai3})$ and $\delta\Theta^c$ through $(\ref{fui})$ and $(\ref{sei23})$, to obtain a second order differential equation for the evolution of the matter contrast,
\begin{equation}\label{bo}
\ddot{\delta}_m^c+\bigg[\frac{2Q}{3\rho_m}+2H+\bigg(A-\frac{\dot K}{K}\bigg)\bigg]\dot{\delta}_m^c+\bigg[\frac{d}{dt}\bigg(\frac{2Q}{3\rho_m}\bigg)+2H\bigg(\frac{2Q}{3\rho_m}\bigg)-\frac{1}{2}\rho_m K+\frac{2Q}{3\rho_m}\bigg(A-\frac{\dot K}{K}\bigg)\bigg]\delta_m^c=0.
\end{equation}
We see that differences arises as compared to the geodesic scenario $(\ref{sin})$, namely, a reduction by a factor $2/3$ in the creation/annihilation rate and a change in the evolution of the dark matter contrast through the scale-dependent function $K$. Furthermore, a new function $A-\frac{\dot K}{K}$ appears in the coefficients of $\dot\delta_m$ and $\delta_m$. The standard $\Lambda$CDM model is recovered if we choose $\alpha=0$. For the case of constant matter creation, which corresponds to the value of $\alpha=-0.5$, the results presented in \cite{W} are recovered.

To estimate the importance of vacuum energy perturbations relative to matter as given by expression $(\ref{fui})$, we start by looking for their values in the deeper matter dominated phase ($z \gg 1$). Since $H\gg Q/\rho_m$ at high redshifts, from $(\ref{poise})$ we have $K \approx 1$, and hence the density contrast is proportional to the scale factor, $\delta_m\propto a$, resulting in the standard growth rate $f=1$. So, the expression $(\ref{fui})$ assumes the scale-independent form
\begin{equation}
\frac{\delta\rho_{V}^c}{\delta\rho_m^c}\approx \frac{2Q}{3\rho_m}\Omega_{m0}^{-\frac{1}{2(1+\alpha)}}z^{-3/2}.
\end{equation}
This ratio is very tiny and depends essentially on the interaction rate and the present value of the matter density. For comparison purposes, if we assume the model with $\alpha = -0.5$ and $\Omega_{m0}=0.45$, corresponding to a constant interaction rate, we found $\frac{\delta\rho_{V}^c}{\delta\rho_m^c}\sim 10^{-5}$ at $z_i=1000$. On the other hand, at the same redshift, for $\alpha = -0.1$ and $\Omega_{m0}=0.3$ we have $\frac{\delta\rho_{V}^c}{\delta\rho_m^c}\sim 10^{-9}$, which is different in several orders of magnitude. 

\begin{figure}
\centerline{\includegraphics[height=5.5cm]{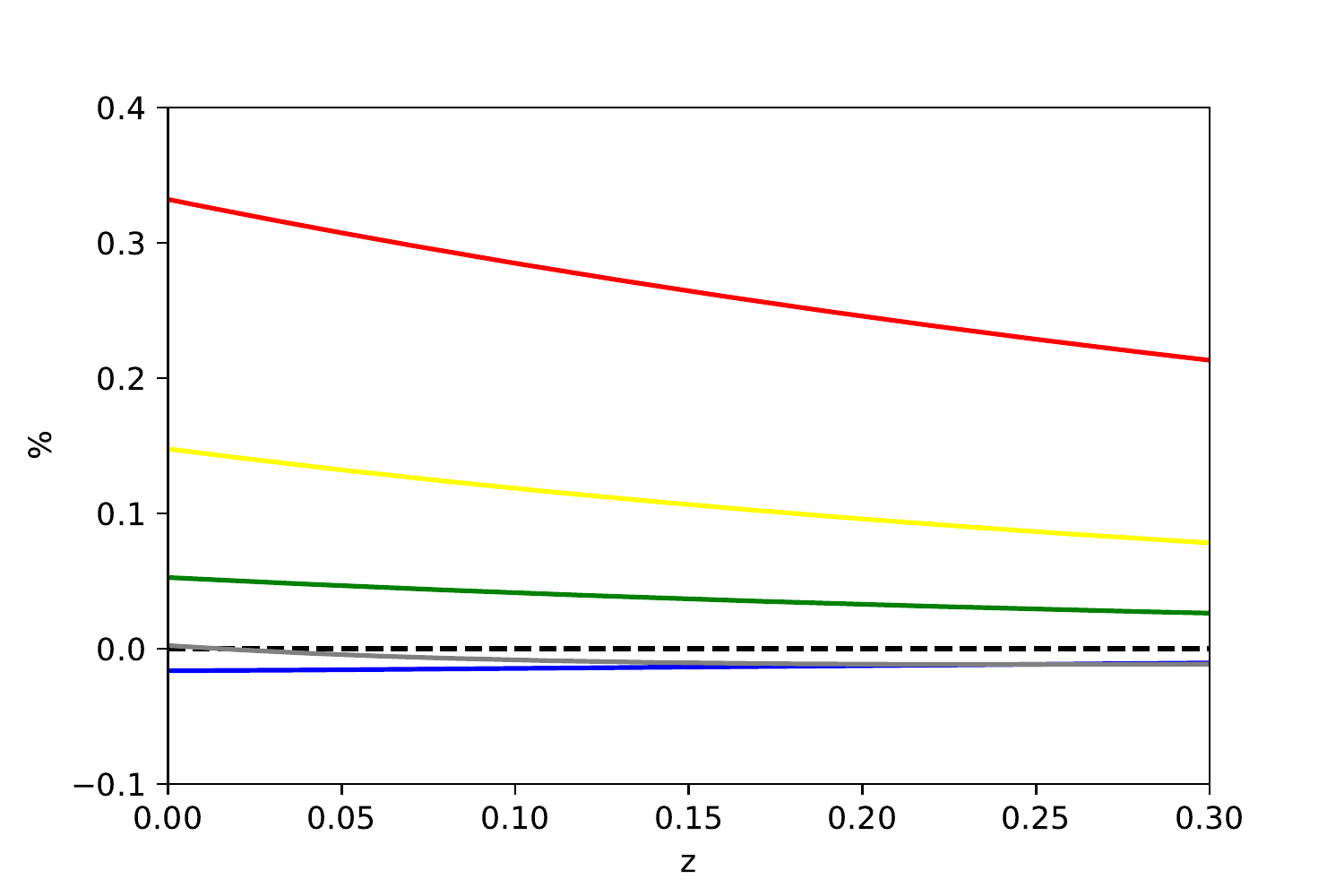}}
\caption{Relative difference between the matter contrasts for the scale $k=0.2$ and for the scale $k=0.01$, as a function of the redshift for different values of the gCg parameter.}
\end{figure}

\begin{figure}
\centerline{\includegraphics[height=5cm]{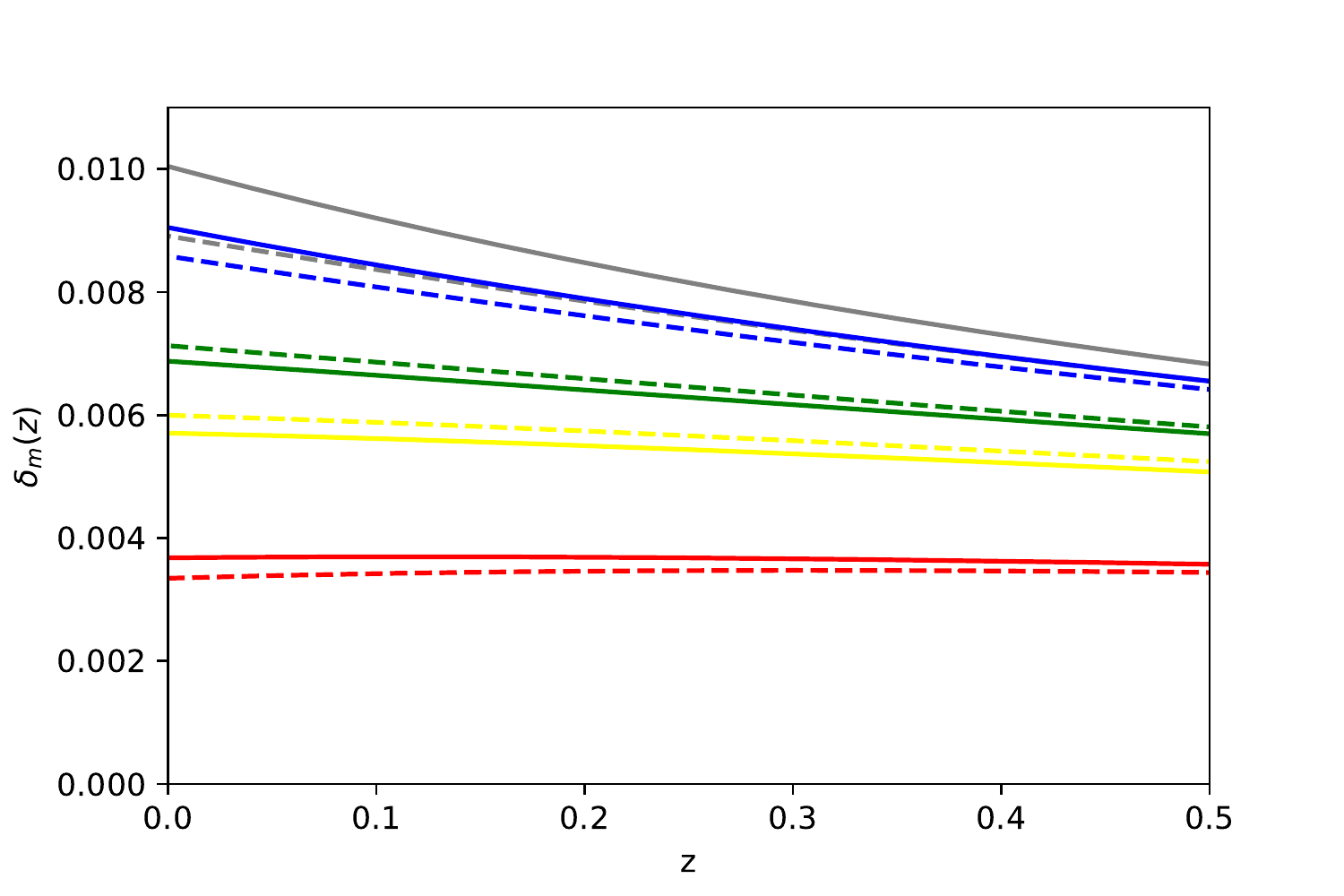} \hspace{.05in} \includegraphics[height=5cm]{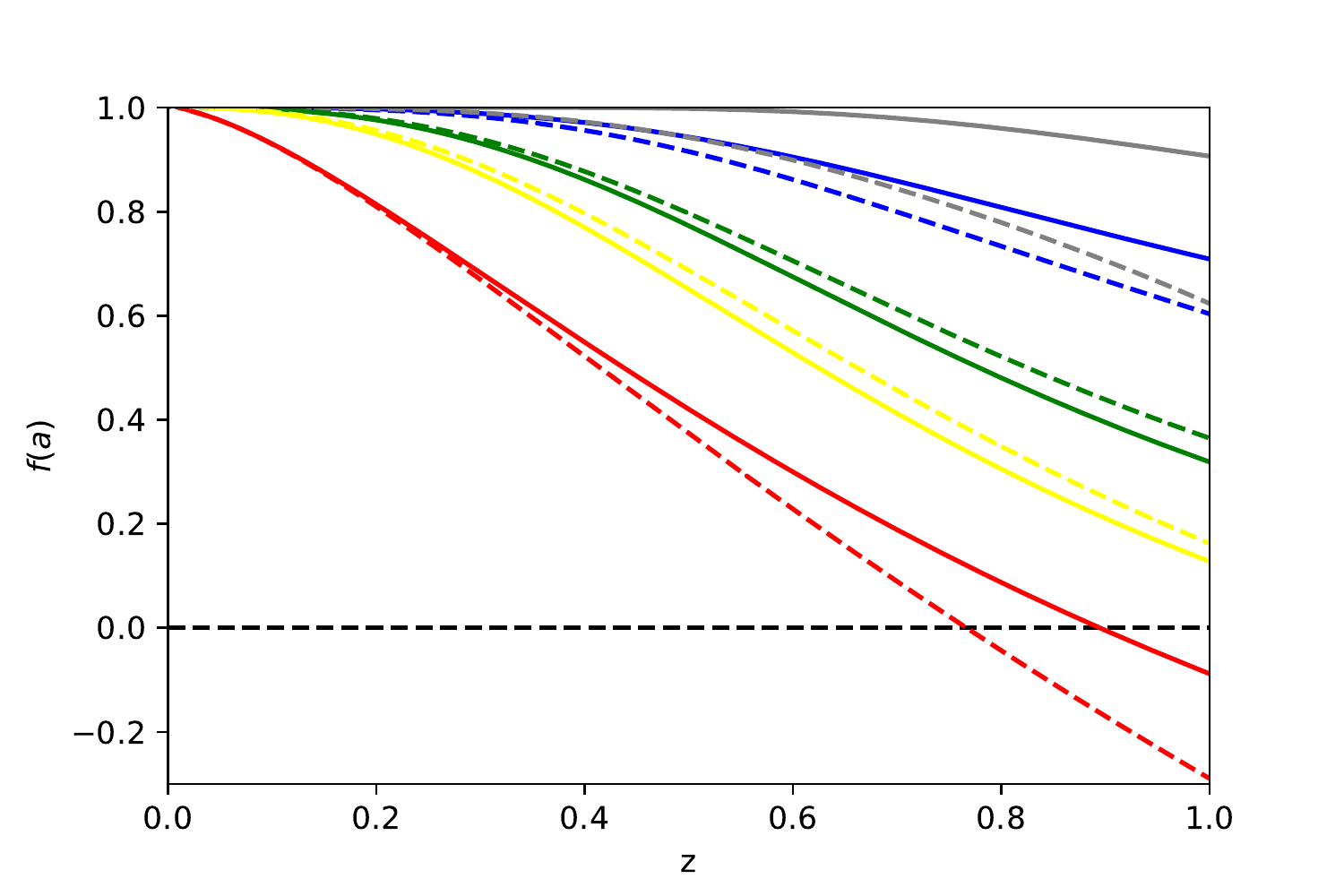}}
\caption{Matter contrast (left panel) and growth rate (right panel) for the following models: $\alpha =-0.5$ (red), $\alpha =-0.2$ (yellow), $\alpha =-0.1$ (green), $\alpha =0.1$ (blue) and $\alpha =0.2$ (gray). We have used the value $\Omega_{m0}=0.45$ for the model with $\alpha=-0.5$. For all the other models we have used $\Omega_{m0}=0.3$. Solid curves correspond to the geodesic and dotted curves to the non geodesic.}
\end{figure}

On the other hand, at late times the vacuum perturbations depend on the scale. The observational data of the linear power spectrum lie in the comoving wave number range $0.01$ Mpc$^{-1}<k<0.2$ Mpc$^{-1}$. In this range, taking the gCg parameter in the interval $-0.5 \leq\alpha\leq 0.5$, we find  the ratio between the vacuum and dark matter perturbations at $z=0$ in the interval
\begin{equation}
10^{-6} < \frac{\delta\rho_{V}^c}{\delta\rho_m^c}< 10^{-4},
\end{equation}
where the upper value corresponds to $k=0.01$, and the lower value to $k=0.2$. Therefore, vacuum perturbations are strongly suppressed inside the Hubble horizon respect to matter perturbations for the class of interacting models considered here. The smaller the scale value, the stronger the suppression. 
These results are shown in Fig. 1, where we plot the ratio between matter perturbations for the scale $k$=0.2 and for the scale $k$=0.01 as a function of the redshift, which is less than $0.4\%$ for $|\alpha| < 0.5$.

To obtain the evolution of $f$ we need to solve the second order equation $(\ref{sin})$ and {(\ref{bo})} by fixing the same initial amplitude $\delta_m(z_i)$ for all models at $z_i=1000$ such that in the matter dominated era we have the standard value $f(z_i)=1$, since the matter contrast is proportional to the scale factor, $\delta_m \propto a$. 
 Fig. 2 shows the matter density contrast and growth rate when we use the geodesic (solid curves) and non geodesic (dotted curves) scenarios. The differences increase for large values of $|\alpha|$. We see an enhancement in the curves as compared to the geodesic for $\alpha = -0.1, -0.2$, and a suppression for $\alpha =0.1, 0.2$. For the case $\alpha = -0.5$, corresponding to matter creation at constant rate, a large suppression appears in the matter growth rate. Therefore, the correction terms introduced in equation $(\ref{bo})$ by the vacuum inhomogeneities should be taken into account when we analyse the growth rate evolution.

\subsection{Newtonian gauge}

Let us show that another way to determine the evolution of the growth rate is to calculate the scalar expansion up to first order, through the covariant derivative of the 4-velocity. In the Newtonian gauge it results in \cite{W. Luciano}
\begin{equation}\label{W1}
\delta\Theta=-\frac{3}{a}(\psi'+\mathcal H \phi)+\frac{\nabla^2v}{a},
\end{equation}
where the prime represents a derivative with respect to the conformal time. We can show that the above expression satisfies the Raychaudhury equation $(\ref{eq2})$, which we have used to find the second order differential equation ($\ref{bo}$). To do that, write the first term of $(\ref{eq2})$ as
\begin{equation}\label{ya}
\delta(\Theta_{,\mu}u^{\mu})=\frac{1}{a}(\delta\Theta'-\Theta'\phi),
\end{equation}
where $u^0=1/a$, $\delta u^0=-\phi/a$ and $\mathcal H=aH$ in the conformal time. The derivative of $(\ref{W1})$ is given by
\begin{equation}\label{ra}
\delta\Theta'=-\frac{3}{a}(\psi''+\mathcal H\phi'+\mathcal H'\phi)+\frac{\nabla^2v'}{a}-\mathcal H\delta\Theta.
\end{equation}
Now consider the set of Einstein's equations in the Newtonian gauge \cite{MH},
\begin{equation}\label{hi2}
-\nabla^2\psi+3\mathcal H(\psi'+\mathcal H\phi)=-\frac{a^2}{2}(\delta\rho_m+\delta\rho_{V}),
\end{equation}
\begin{equation}\label{hi1}
\psi''+2\mathcal H\psi'+\mathcal H\phi'+(2\mathcal H'+\mathcal H^2)\phi=-\frac{a^2}{2}\delta\rho_{V}+\frac{a^2}{3}\nabla^2\pi,
\end{equation}
\begin{equation}\label{hi3}
\psi'+\mathcal H \phi=-\frac{a^2}{2}\rho_mv,
\end{equation}
\begin{equation}\label{hi4}
\psi-\phi=a^2\pi,
\end{equation}
where $\pi$ is the anisotropic stress perturbation.
Using $(\ref{hi1})$, $(\ref{hi2})$ and $(\ref{W1})$ in Eq. $(\ref{ra})$ it is not difficult to show that $(\ref{ya})$ is given by
\begin{equation}
\delta(\Theta_{,\mu}u^{\mu})+\frac{2}{3}\Theta\delta\Theta-\frac{\nabla^2}{a^2}(v'+\psi)+\frac{1}{2}(\delta\rho_m-2\delta\rho_V)=\nabla^2 \bigg(\frac{\mathcal H}{a^2}v -\pi\bigg).
\end{equation} 
Finally, we use Eq. $(\ref{hi4})$, make the substitution $v\rightarrow v/a$ and change from conformal time to cosmological time to obtain the Raychaudhury equation
\begin{equation}
\delta(\Theta_{,\mu}u^{\mu})+\frac{2}{3}\Theta\delta\Theta-\frac{\nabla^2}{a^2}(\dot v+\phi)+\frac{1}{2}(\delta\rho_m-2\delta\rho_V)=0.
\end{equation}

We can also obtain the evolution of coupled equations for total matter density contrast $\delta_m$, matter velocity defined by $\theta_m=-k^2v$ and gravitational potential $\phi$ in Newtonian gauge. We consider the anisotropic stress perturbations equal to zero, which implies $\phi=\psi$.

In the limit of small scales $k \gg aH$, after some calculations using (\ref{W1}), (\ref{open45}), (\ref{paises4}), (\ref{sei23}) and (\ref{sei}) into (\ref{mn}), it is straightforward to show that the energy balance assumes the form
\begin{equation}
\delta'_{m}+\theta_{m}+\frac{aQ}{\rho_{m}} \delta_{m}=\frac{aQ}{3\rho_m}\delta_m+\frac{4Q^2H^2}{9\rho_{m}^4}\bigg(\frac{k}{aH}\bigg)^2\theta_m.
\end{equation}
 In the sub-horizon limit $k\gg aH$, the Poisson equation (\ref{hi2}) is given by
\begin{equation}\label{poiss}
-k^2\phi=\frac{a^2}{2}\rho_m\delta_m,
\end{equation}
where we neglected the vacuum energy perturbation $\delta\rho_{V}$ as compared with matter density perturbation $\delta\rho_m$ and the second term on the right of equation.
Finally, it is easy to show that the momentum balance (\ref{sei}) becomes
\begin{equation}
\theta'_{m}+aH\theta_{m}-k^2\phi= -\frac{2aQH^2}{3\rho_{m}^2}\bigg(\frac{k}{aH}\bigg)^2\theta_m.
\end{equation}

For the geodesic case in the sub-horizon limit, the resulting equations are given in the reference \cite{Benetti:2019lxu}.

\subsection{Perturbation of baryons}

In this section the perturbation of baryons will be included in the dynamics to account the effect of Redshift-Space Distortion (RSD) on large scale structures (LSS) data. The assumption is that baryons reflect the motion of galaxies, since what we directly observe is their emitted light. 

To derive the evolution equation for the baryonic contrast $\delta_b$ in the Newtonian gauge we assume that vacuum does not interact with baryons. Thus, the resulting continuity and Euler equations in the sub-horizon limit, are given, respectively, by \cite {rds}
\begin{equation}
\delta'_b+\theta_b=0,
\end{equation}
\begin{equation}\label{rai1}
\theta'_{b}+aH\theta_{b}-k^2\phi=0.
\end{equation}

For $\Lambda$CDM model where the energy transfer is absent ($Q$=0) $\delta_m$ and $\delta_b$ have exactly the same evolution, however the suppression or enhanced in the matter density contrast for models with $\alpha\neq 0$ affect the baryon density contrast through Poisson equation ($\ref{poiss}$). 

Let us define the linear matter power spectrum in Fourier space at a redshift $z$, as
\begin{equation}\label{sui}
P_{m,b}(k,z)=P(k,z_i)(\delta_{m,b}(z))^2,
\end{equation}
where the suffices $m$ and $b$ refer to the baryon and total matter and $P(k,z_i)$ is the matter power spectrum defined at matter dominated era.

The amplitude of baryon and total matter density fluctuation within sphere of comoving radius 8 $h^{-1}$Mpc at redshift $z$ is given by
\begin{equation}\label{pion}
\sigma_{8m,b}(z)=\bigg[\frac{1}{2\pi^2}\int k^2dk W(kR)P(k,z_i)\bigg]^{1/2}\delta_{m,b}(z),
\end{equation}
with the window function
\begin{equation}\label{pion02}
W(kR)=\bigg[\frac{sin(kR)}{(kR)^3}-\frac{cos(kR)}{(kR)^2}\bigg].
\end{equation}
Now, using ($\ref{pion}$) we can take the relation for each component
\begin{equation}\label{sigmas_relation01}
\sigma_{8m,b}(z)=\sigma_{8m,b}(0)\frac{\delta_{m,b}(z)}{\delta_{m,b}(0)}.
\end{equation}
Similarly, we can derive $\sigma_{8m}$ from $\sigma_{8b}$ through the relation
\begin{equation}\label{relation04}
\sigma_{8m}(0)=\sigma_{8b}(0)\frac{\delta_{m}(0)}{\delta_{b}(0)}.
\end{equation}
For $\Lambda$CDM model we have $\sigma_{8m}(0)=\sigma_{8b}(0)$ due to the absence of interactions. However, for interaction models the today's amplitude of baryons and total matter density fluctuations are, in fact, different by the above relationship.
From ($\ref{sui}$) and ($\ref{relation04}$) it is possible to find the baryon power spectrum in terms of the total matter power spectrum at $z$=0
\begin{equation}\label{sui00}
P_{b}(k,0)=P_m(k,0)\bigg[\frac{\sigma_{8b}(0)}{\sigma_{8m}(0)}\bigg]^2.
\end{equation}
Now we can compute the observed galaxy power spectrum in redshift space as
\begin{equation}
P_g(k,z)=[b_1\sigma_{8m}(z)+\mu^2_{k}f_b(z)\sigma_{8b}(z)]^2\frac{P_m(k,0)}{\sigma_{8m}^2(z)},
\end{equation}
onde $b_1$ is the linear bias and $\mu_k$ is the cosine of the angle between the line of sight and the wavevector $\hat k$. Regarding to the RSD, the important quantity 
is the bias independent combination  $f_b(z)\sigma_{8b}(z)$, where $f_b$ is the baryon growth rate and $\sigma_{8b}$ represents the amplitude of baryon density fluctuation within sphere of comoving radius 8 Mpc given by $(\ref{sigmas_relation01})$.

In Fig. 3 we plot the evolution of $f_b(z)\sigma_{8b}(z)$ on the left panel for $\alpha=0$, $\alpha=0.2$ and $\alpha=-0.5$ fixing $\sigma_{8b}(0)=0.83$. The values of the total density parameters are chosen to be $\Omega_{m0}=0.45$ for $\alpha=-0.5$ and $\Omega_{m0}=0.3$ for the other two models. In the right panel, we show the relative difference between homogeneous and inhomogeneous vacuum ansatz for the expected combination of RSD for baryons.
\begin{figure}
\centerline{\includegraphics[height=5cm]{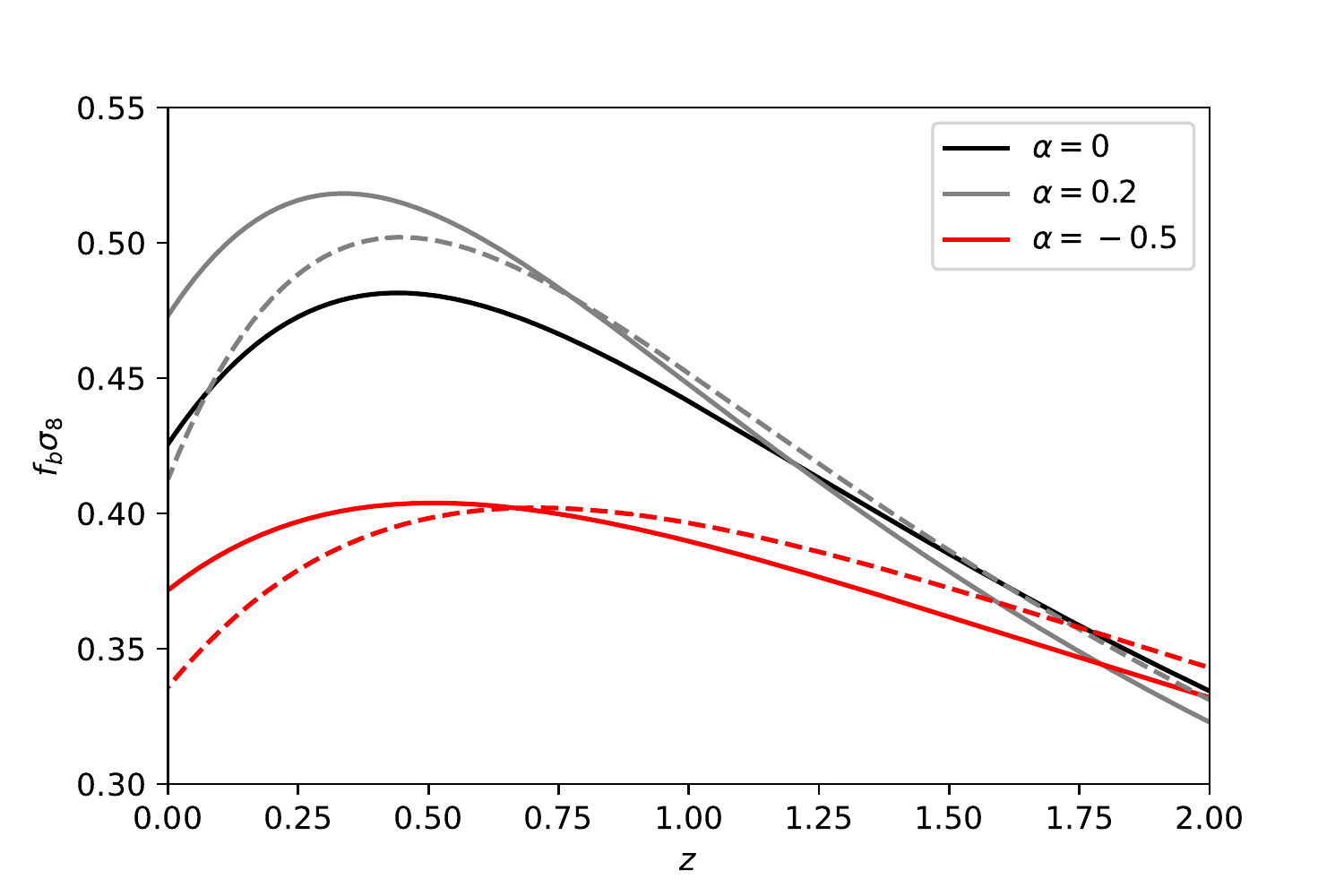} \hspace{.25cm}  \includegraphics[height=5cm]{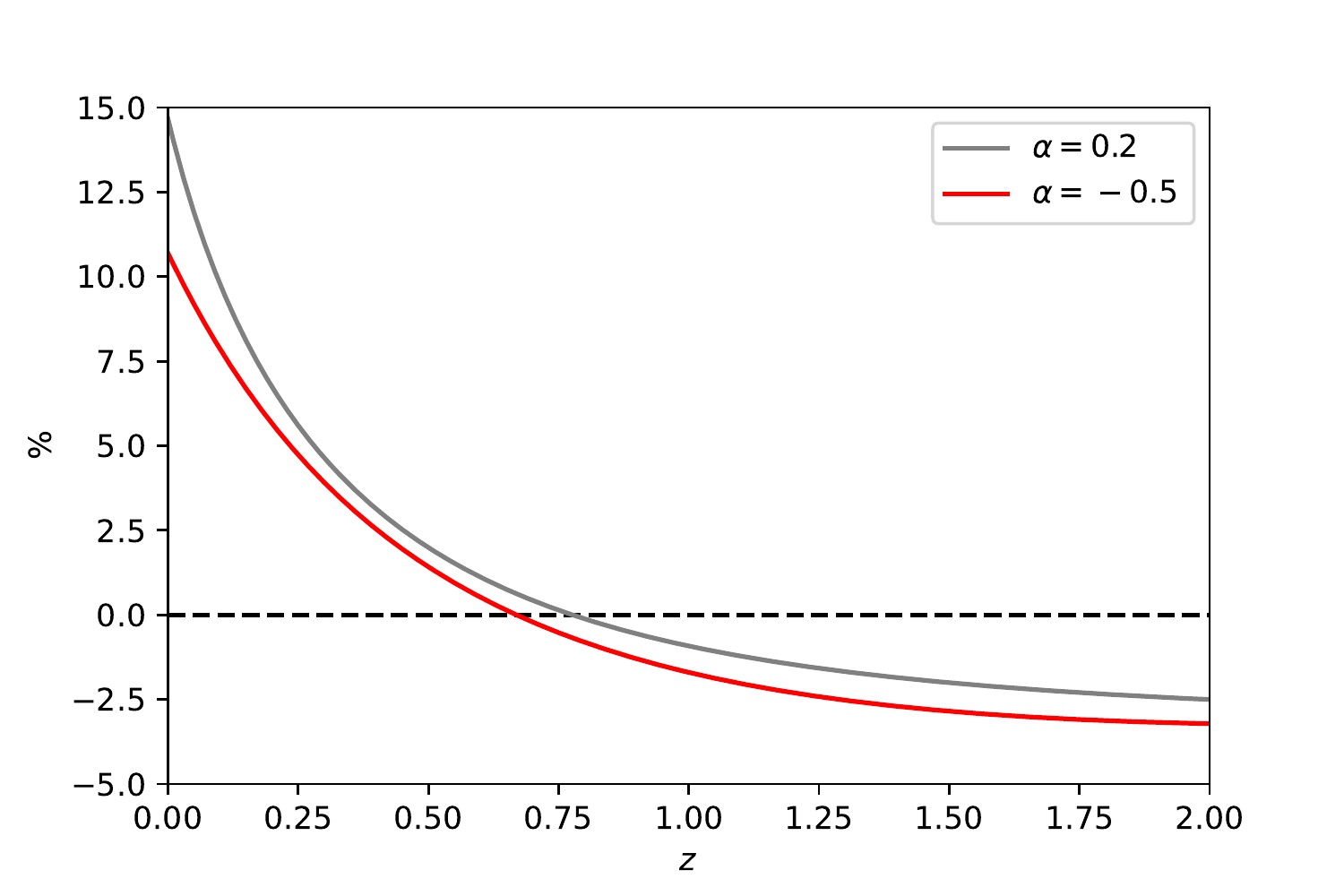}}
\caption{Left panel: The evolution of combination $f_b(z)\sigma_{8b}(z)$ for $\Lambda$CDM (black curve), for $\alpha =0.2$ (gray curve) and for $\alpha =-0.5$ (red curve). Solid curves correspond to the geodesic model and dotted curves to inhomogeneous vacuum. Right panel: The percent relative difference between homogeneous and inhomogeneous vacuum ansatz.}
\end{figure}

\section{Observational tests}

Once we have the description of both homogeneous and inhomogeneous vacuum models, equations (\ref{sin}) and (\ref{bo}), it is possible to construct the quantity $f_b(z)\sigma_{8b}(z)$ and perform observational tests against available data. Our goal here is to compare the constraints obtained with redshift space distortion data alone, and also to check the impact of this analysis when combining with independent data with bigger sample and more controlled errors. For that, we used type Ia supernovae samples to combine with  $f_b\sigma_{8b}$ data set. In this section we present the data used in our analysis, the applied statistics and some results.

\subsection{Data set and Method}

In our analysis, we choose to use a joint data set combining several probes, such as Redshift Space Distortion data, Supernovae Ia samples and Cosmic Microwave Background likelihoods, in order to cover evolution of the universe at different scales. 

About the RSD probe, we considered $(f\sigma_8)'$ data presented in Tab. IV of Ref. \cite{skara_tension_2020}, containing each fiducial cosmology on the last column, and the prime refers to the value in the fiducial cosmology. Taking into account corrections due to Alcock-Paczynski effect, the value $(f\sigma_8)$ for the cosmology of interest is obtained making

\begin{align}\label{eq:EquaQueTemOQ}
f \sigma_8 \sim q (f\sigma_8)', 
\end{align}
where
\begin{align}
q = \frac{H(z|s)D_A(z|s)}{H'(z|s')D'_A(z|s')},
\end{align}
with $s'$ and $s$ representing sets of cosmological parameters in both fiducial and tested models.
In order to not having to evaluate errors and covariance matrix for the models of interest, avoiding error propagation in this process, we built the likelihood for RSD data proportional to $\exp(-\chi^2/2)$, where 
\begin{align}
\chi^2 = \sum_{ij} V'_i(C'^{-1}_{ij})V'_j,
\end{align}
and, 
\begin{align}
V'_i \equiv (f \sigma_{8})'_i - \frac{f\sigma_8(z_i|s)}{q(z_i|s,s')}.
\end{align}

Only WiggleZ data present correlations, translated into the following covariance matrix,
\begin{align}
C_{ij}'^{\text{WiggleZ}}=10^{-3} \left(\begin{array}{ccc}
6.400 & 2.570 & 0.000\\
2.570 & 3.969 & 2.540\\
0.000 & 2.540 & 5.184\\
\end{array}\right).
\end{align} 

At the same time, we consider both Supernovae Ia by the Joint Light-curve (JLA) sample~\cite{Betoule:2014frx}, and Pantheon compilation~\cite{Scolnic:2017caz}.
The first is a sample of 740 data points of the Supernova Legacy Survey (SNLS) and Sloan Digital SkySurvey (SDSS), covering the redshift range $0.01< z <1.3$. This likelihood is marginalized over two nuisance parameters, with the advantage of allowing the light-curve recalibration with the model under consideration, which is an important issue when testing alternative cosmologies~\cite{Taddei:2016iku,Benetti:2019lxu}. 
Instead, the Pantheon compilation cover the range $0.01< z <2.3$ with 1048 SNe Ia, combining the subset of 276 newPan-STARRS1 SNe Ia with useful distance estimates of SNIa from SNLS, SDSS, low-z and Hubble space telescope (HST) samples. 

Finally, we also take into account CMB measurements, through the Planck (2018) data~\cite{Aghanim:2019ame}, using ``TT,TE,EE+lowE" data by combination of temperature power spectra and cross correlation TE and EE over the range $\ell \in [30, 2508]$, the low-$\ell$ temperature Commander likelihood, and the low-$\ell$ SimAll EE likelihood. We refer to this data set as ``Planck18".

Our analysis are divided into two distinct methodologies. The first is a RSD and SNe data joint analysis, building chains with the Python MultiNest algorithm \cite{feroz_multinest_2009,Buchner}. We set $1500$ livepoints and a sampling efficiency of $0.5$, giving both parameter and evidence estimations. Those PyMultiNest parameters were chosen to guarantee stability in the exploration of the space parameter achieving and acceptable convergence.

\begin{table}
	\centering
	\begin{tabular}{c | c }
	\hline
	\hline
	Parameter			&  Priors				 	\\ 
	\hline 
	$H_0$ (km/s/Mpc) & ${\cal{G}}[74.03,1.42]$		\\ 
	$\Omega_bh^2$	&  ${\cal{G}}[0.02166,0.00026]$	\\ 
	$\Omega_{c}h^2$	& $\mathcal{U}[0.01,0.25]$	\\ 
	$\alpha$	   	&  $\mathcal{U}[-0.99,1.00]$		\\ 
	$\sigma_{8}$ & $\mathcal{U}[0.4,1.2]$  \\

	\hline 
	$\alpha_s$	   	&  $\mathcal{U}[0.0,1.0]$	\\ 
	$\beta_s$	   	&  $\mathcal{U}[0.0,4.0]$	\\ 
	$\mathcal{M}_B$	   	&  $\mathcal{U}[-22,-16]$		\\ 
	$\Delta_M$	   	&  $\mathcal{U}[-1.0,1.0]$			\\ 
	\hline 
	\hline
	\end{tabular}
	\caption{Adopted priors for cosmological and supernovae (JLA) nuisance free parameters on performed analysis. A Gaussian prior for $H_0$ inspired on Riess \emph{et al.} measurements was chosen, once this parameter is not relevant for RSD-only analysis and for supernovae data fitting, $H_0$ is fully correlated to $\mathcal{M}_B$. $\sigma_{8}$, is a derived parameter.}
	\label{tab:priorstabel}
\end{table}

The priors used in our analysis are summarized in Table \ref{tab:priorstabel},  where Gaussian priors are indicated for local Hubble-{Lema\^itre} parameter, $H_0$ \cite{priorH0_Riess}, and the physical baryon density, $\Omega_{b}h^2$ \cite{prioOmgeabh2_Cooke}, and uniform priors for the physical cold dark matter density,  $\Omega_{c}h^2$, for interaction parameter, $\alpha$, and for $\sigma_{8b}$. Also, uniform priors were adopted for nuisance parameters associate with SNeIa light curve calibration, i.e.  \{$\alpha_s$,$\beta_s$,$\mathcal{M}_B$,$\Delta_M$\} for JLA likelihood, and $\mathcal{M}_B$ for Pantheon likelihood.

The second methodology regards the analysis using the CMB data, i.e when the combined CMB+SNe data set are considered. In this case we modified both the background and perturbation equations, explained in the previous sections, in the numerical Cosmic Linear Anisotropy Solving System (CLASS) code~\cite{Blas:2011rf}, choosing to  work with the Newtonian gauge. In this way, we get the theoretical prediction of cosmological observables, i.e. among others the TT, TE and EE anisotropy CMB power spectra as well as the matter power spectrum. Then, using the Monte Python~\cite{Audren:2012wb} code it is possible to constrain the parameters of the theory by a  Monte Carlo Markov Chains analysis, using the chosen data set. These chains, are then post-processed using RSD data.
In these analysis, we vary the standard six $\Lambda$CDM cosmological parameters in large and flat priors. This means that, besides the $H_0$, $\Omega_{b}h^2$ and $\Omega_{c}h^2$, we also consider the optical depth, $\tau_{reio}$, the primordial scalar amplitude, $\mathcal A_s$, the primordial spectral index, $n_s$, in addition to the interaction parameter, $\alpha$. Also, we consider the CMB nuisance foreground parameters~\cite{Aghanim:2015xee}. Once $\sigma_{8}=\sigma_{8m}$ is given as a derived parameter in MontePython analysis, with the help of (\ref{relation04}) it was possible to also derive $\sigma_{8b}$.

\renewcommand{\arraystretch}{1.4}
\begin{table}
    \centering
\begin{tabular}{c|c|c|c|c}
\hline
               Parameter &     Data                 & \hspace{.7cm} $\Lambda$CDM  \hspace{.7cm}              & Geodesic CG         & non-Geodesic CG     \\
\hline
\multirow{3}{*}{$\alpha$}          & RSD        & 0                         & $-0.011^{+0.695}_{-0.494}$ & $-0.14^{+0.336}_{-0.343}$  \\
               & RSD+JLA     & 0                         & $-0.112^{+0.382}_{-0.316}$ & $-0.111^{+0.289}_{-0.266}$ \\
               & RSD+Pantheon & 0                         & $-0.062^{+0.366}_{-0.298}$ & $-0.053^{+0.249}_{-0.269}$ \\
\hline              
 \multirow{3}{*}{$\Omega_{m0}$}    & RSD          & $0.28^{+0.056}_{-0.051}$   & $0.294^{+0.163}_{-0.156}$   & $0.333^{+0.132}_{-0.097}$   \\
               & RSD+JLA      & $0.286^{+0.042}_{-0.04}$   & $0.313^{+0.092}_{-0.088}$   & $0.314^{+0.079}_{-0.068}$   \\
               & RSD+Pantheon & $0.291^{+0.034}_{-0.032}$  & $0.305\pm0.071$             & $0.305^{+0.059}_{-0.051}$   \\
\hline               
 \multirow{3}{*}{$\sigma_{8b}$} & RSD           & $0.779^{+0.036}_{-0.034}$ & $0.794^{+0.142}_{-0.094}$  & $0.803^{+0.103}_{-0.060}$   \\
               & RSD+JLA       & $0.775^{+0.032}_{-0.031}$ & $0.807^{+0.11}_{-0.08}$    & $0.796^{+0.077}_{-0.051}$  \\
               & RSD+Pantheon  & $0.773\pm0.029$           & $0.792^{+0.096}_{-0.072}$  & $0.785^{+0.069}_{-0.043}$  \\
\hline               
 \multirow{3}{*}{$\sigma_{8}$} & RSD         & $0.779^{+0.036}_{-0.034}$ & $0.805^{+0.582}_{-0.254}$  & $0.69^{+0.143}_{-0.196}$   \\
               & RSD+JLA     & $0.775^{+0.032}_{-0.031}$ & $0.723^{+0.220}_{-0.148}$   & $0.709^{+0.122}_{-0.156}$  \\
               & RSD+Pantheon & $0.773\pm0.029$           & $0.745^{+0.206}_{-0.144}$  & $0.736^{+0.099}_{-0.149}$  \\
\hline
\end{tabular}
    \caption{Results obtained when constraining the discussed models against RSD data, and also the joint analysis with both SNeIa samples. Parameters mean values and $95\%$ credible intervals are presented.}
    \label{tab:results_RSD_SNe}
\end{table}
\renewcommand{\arraystretch}{1}

\begin{figure}
    \centering
    \includegraphics[width=0.4\textwidth]{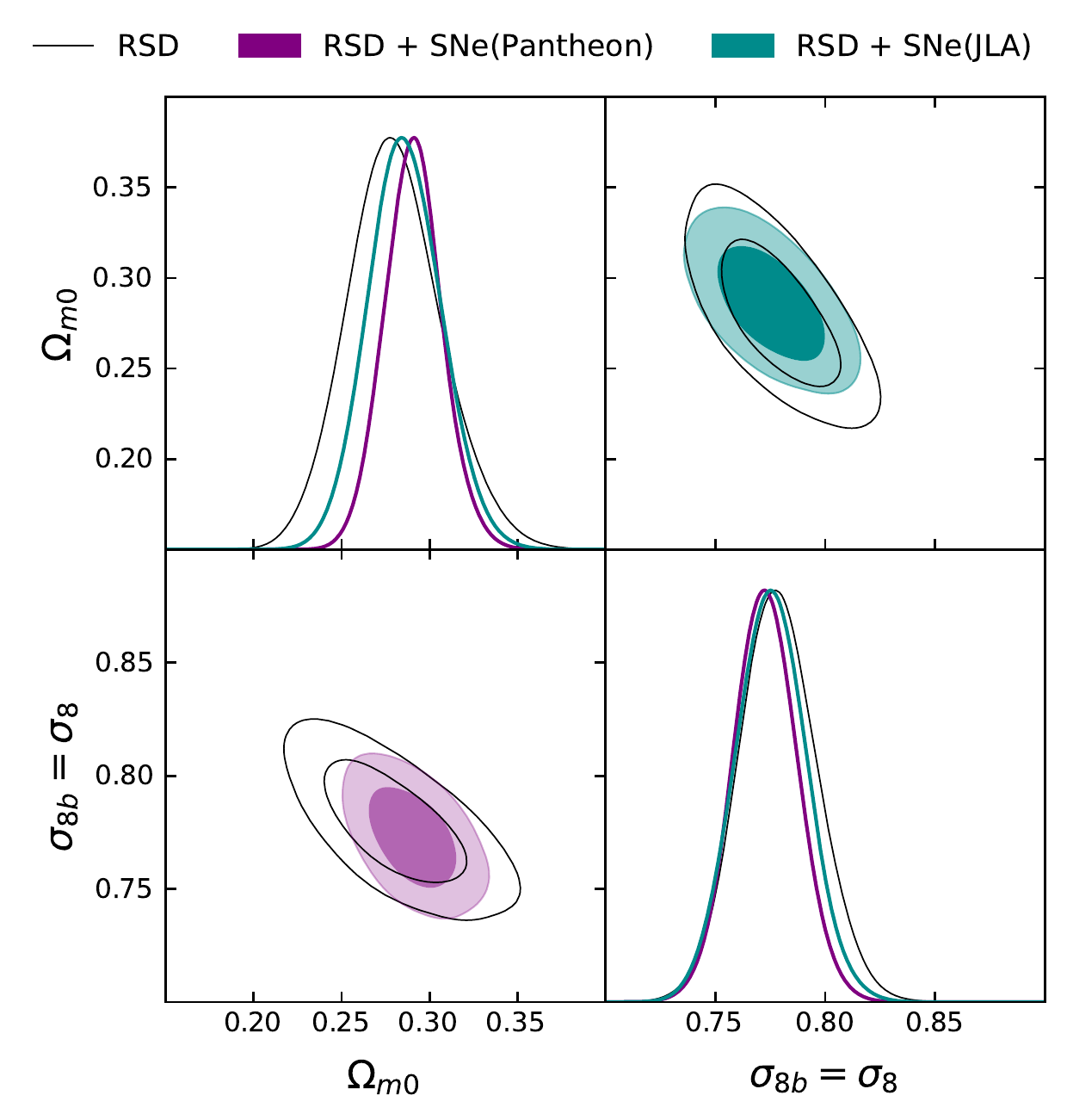}
     \vspace{-0.1in}
    \caption{RSD and RSD+SNeIa posterior distributions and 2-D credible regions for $\Omega_{m0}$ and $\sigma_8$ in the $\Lambda$CDM scenario. RSD results show a good agreement with SNeIa, and the joint analysis return tighter constraints on matter density parameter.}
    \label{fig:standard}
\end{figure}

\begin{figure}

    \centering
    \includegraphics[width=0.488\textwidth]{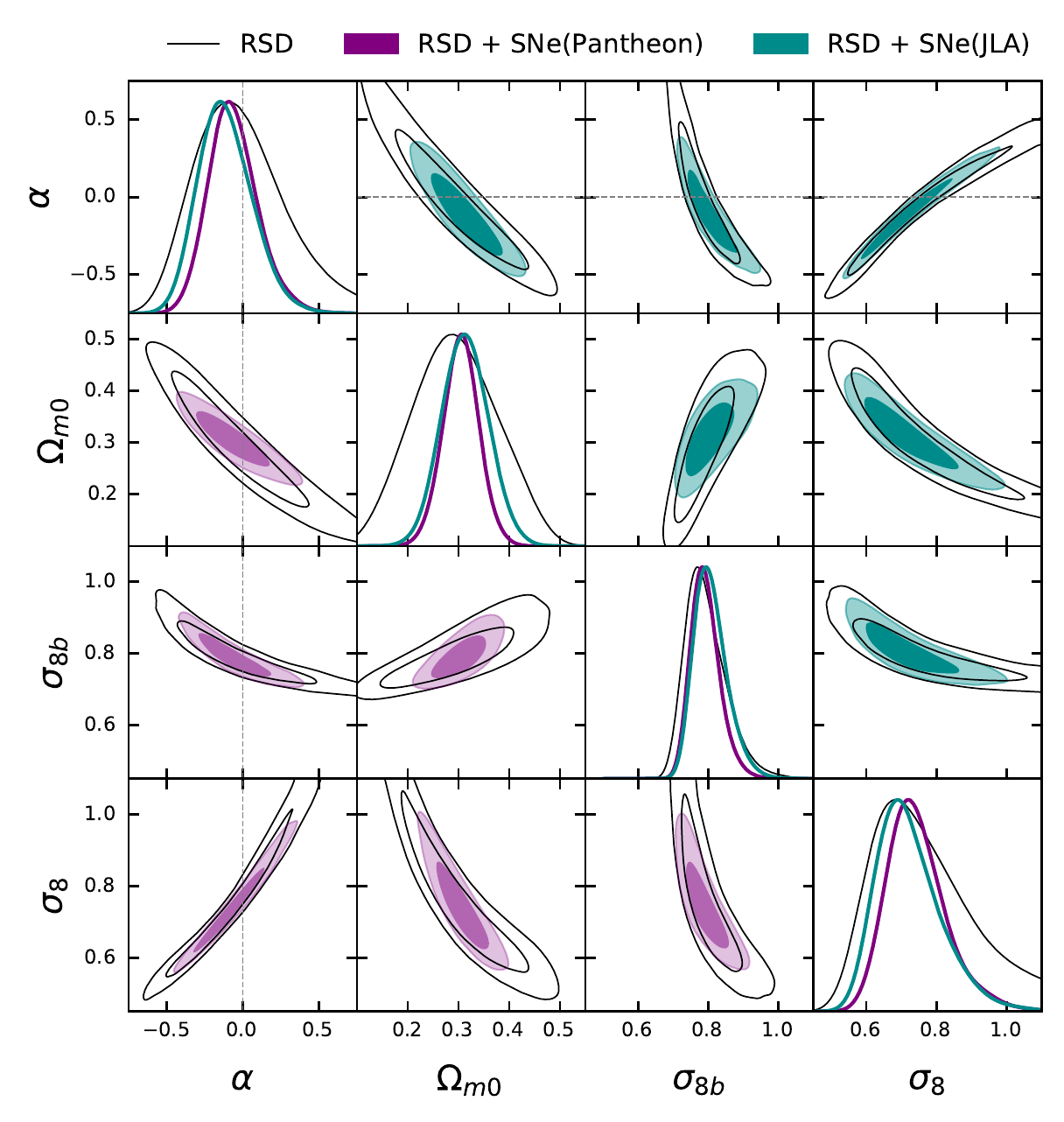}
\hfill
    \includegraphics[width=0.488\textwidth]{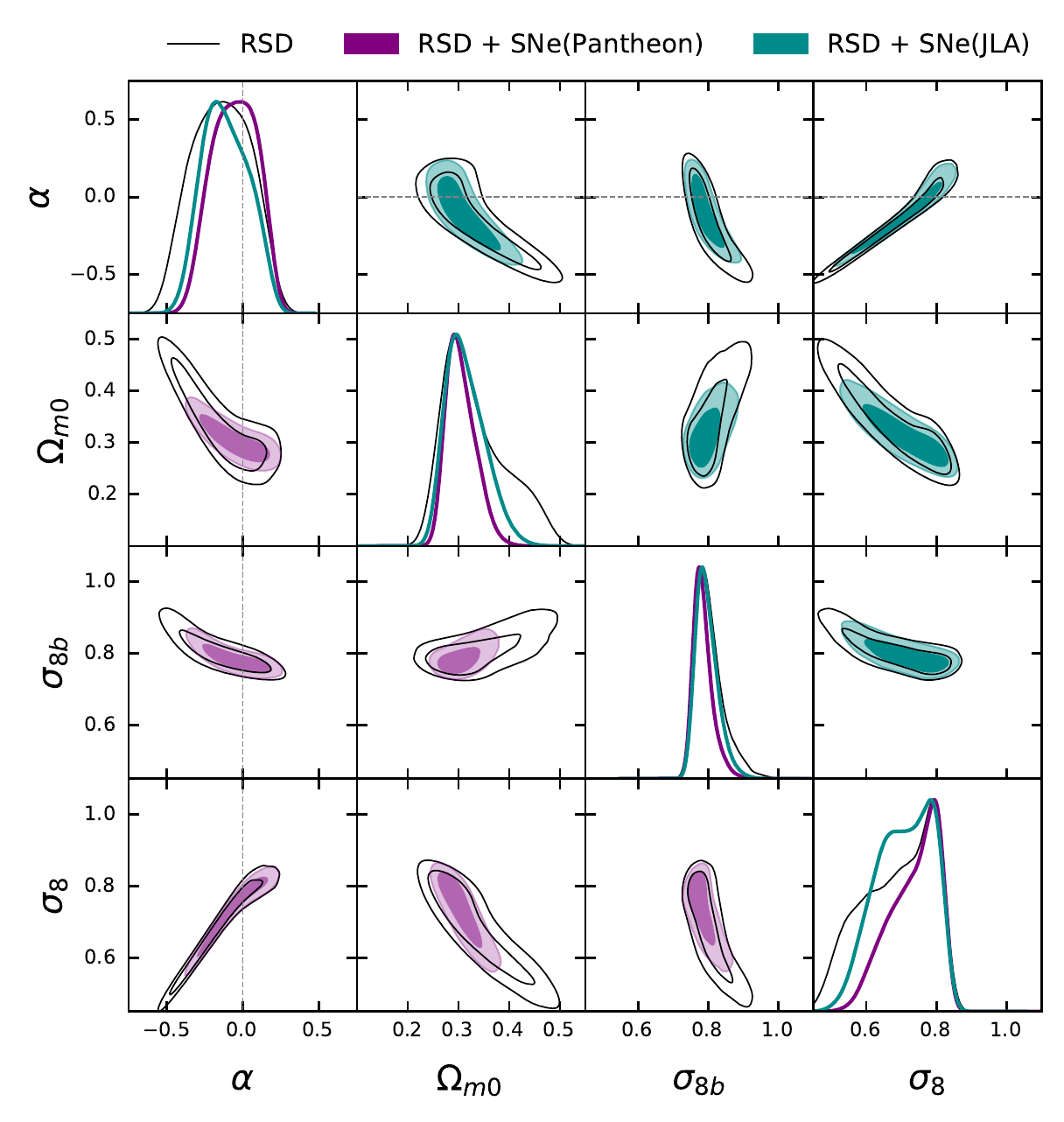}
    \caption{Geodesic Chaplygin gas (left panel) and Non-Geodesic (right panel) parameters plots of 1D PDF and 2D credible regions (1$\sigma$ and 2$\sigma$) as results of statistical analysis with RSD and RSD+SNeIa. Those results are summarized on table \ref{tab:results_RSD_SNe}. }
    \label{fig:rsd_sne}
\end{figure}

\begin{figure}
    \centering
    \includegraphics[width=0.488\textwidth]
    {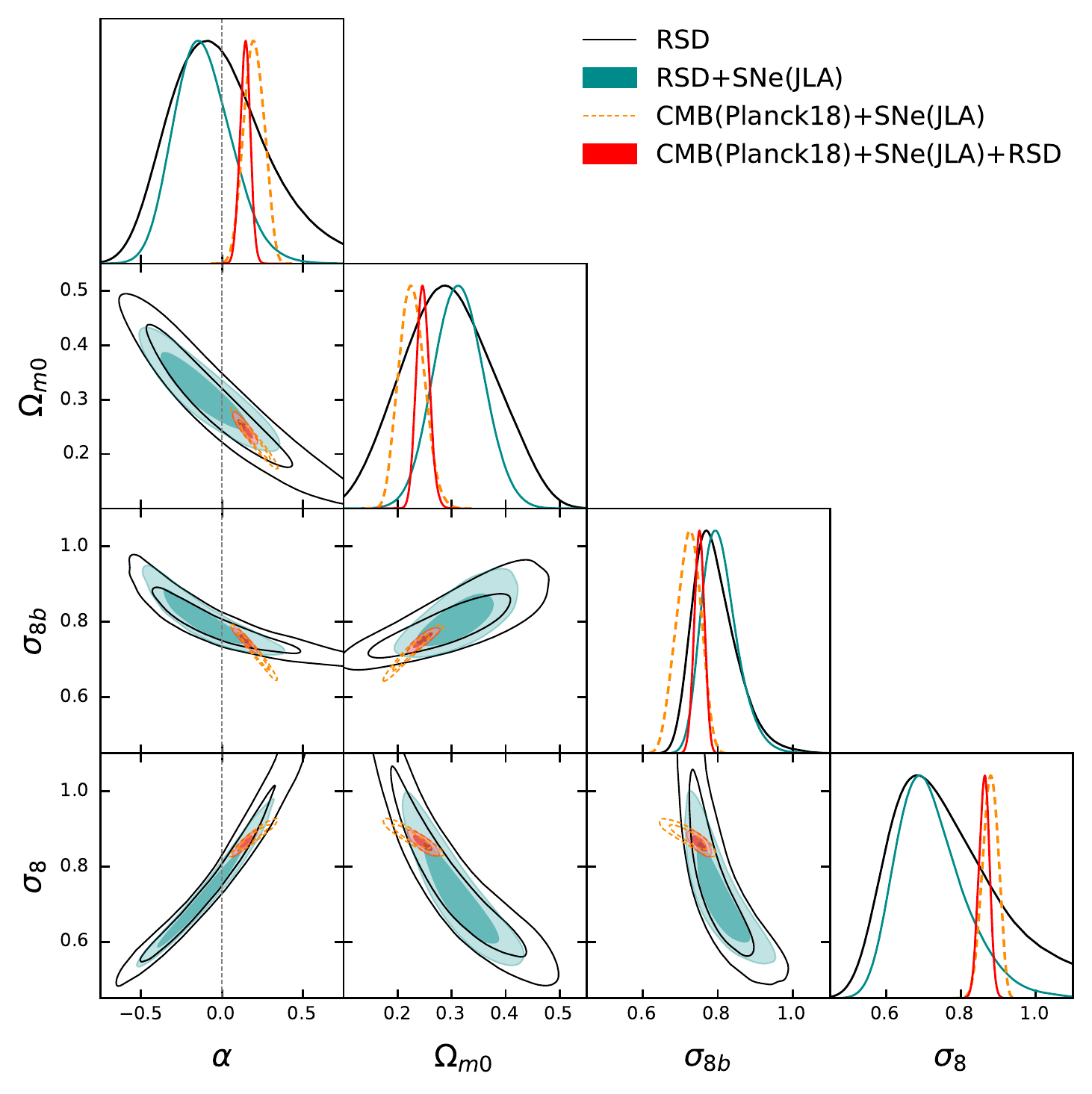}
\hfill
    \includegraphics[width=0.488\textwidth]{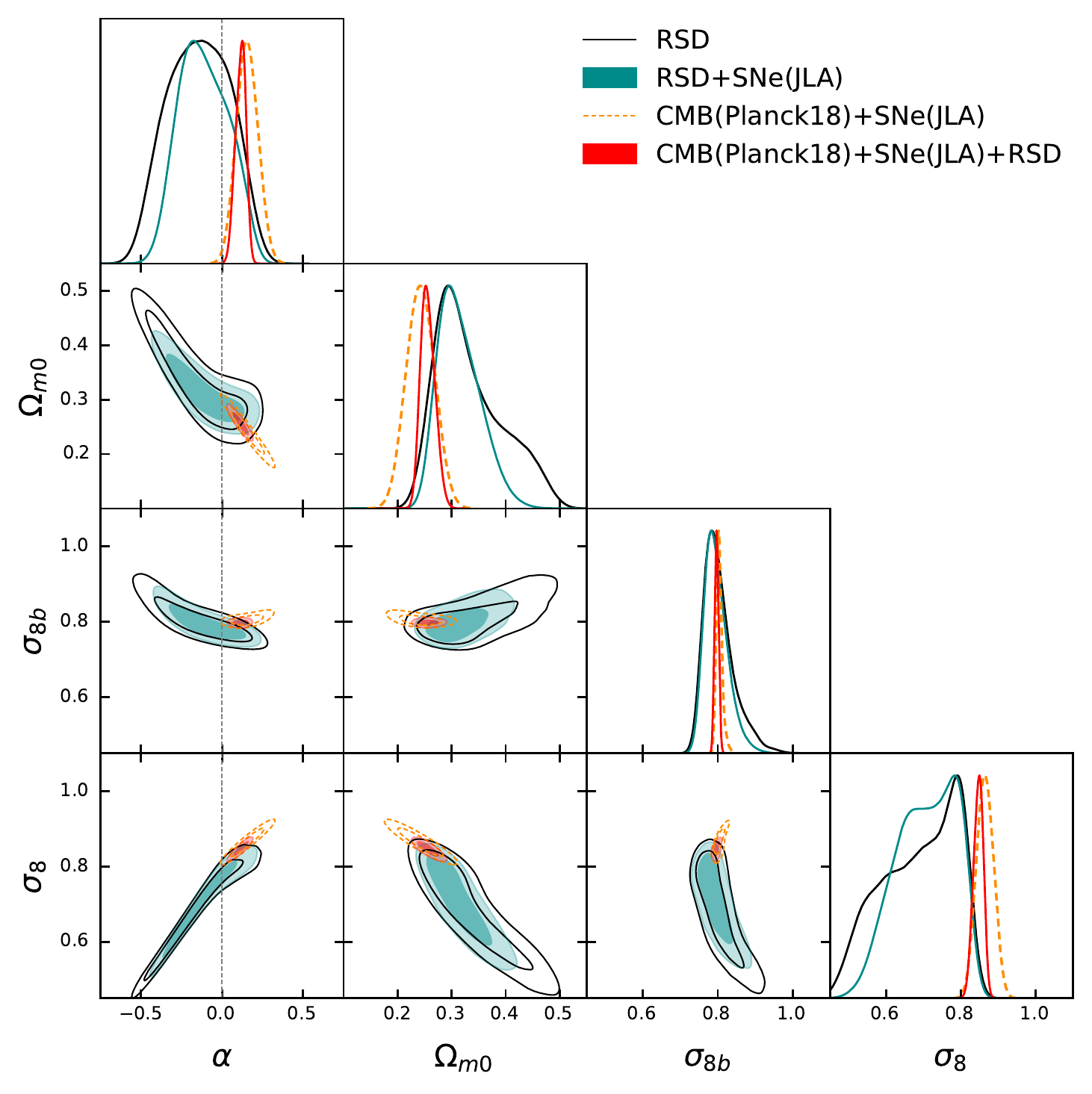}

    \caption{{\bf Left:} Geodesic Chaplygin gas; {\bf Right:} non-geodesic. Results obtained with RSD only and some combinations of datasets. Red curves show the effect of post-processing Planck18+JLA chains with RSD likelihood. Mean and credible intervals are presented at table \ref{tab:final_table}. }
    \label{fig:full}
\end{figure}

\begin{figure}
    \centering
    \hspace{-.9cm}
    \includegraphics[width=0.85\textwidth]
    {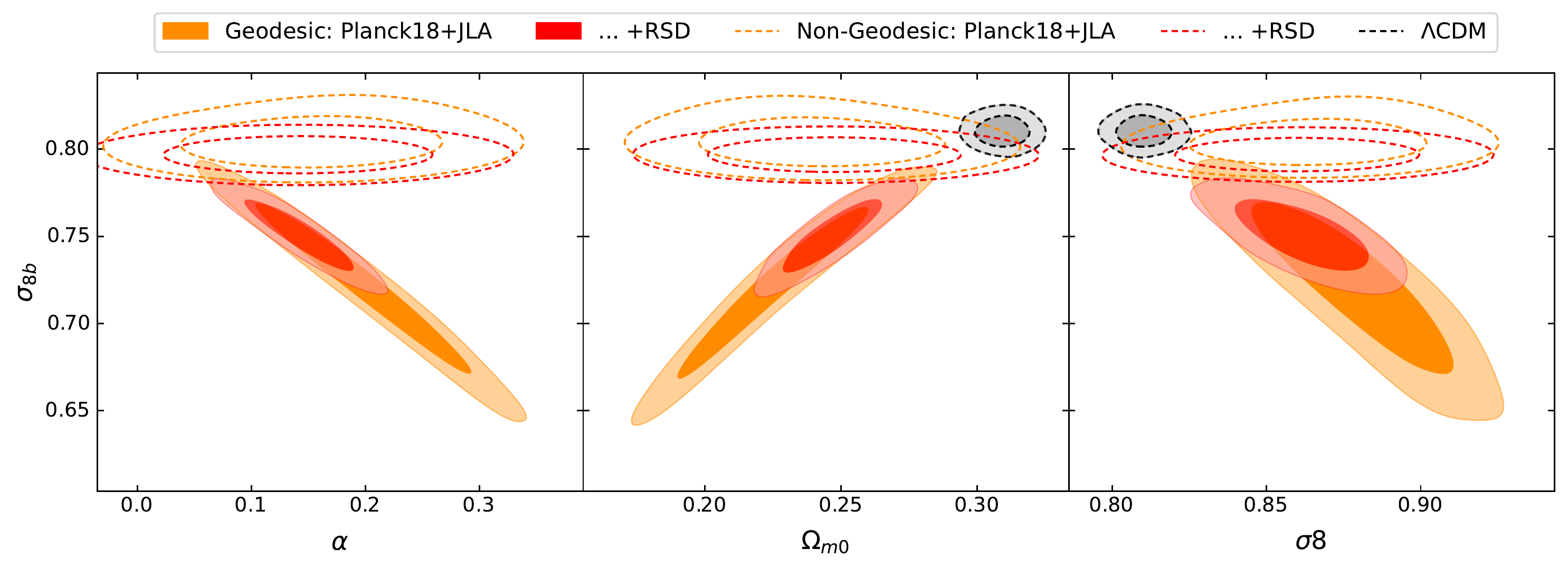}
\hfill
    \includegraphics[width=0.8\textwidth]{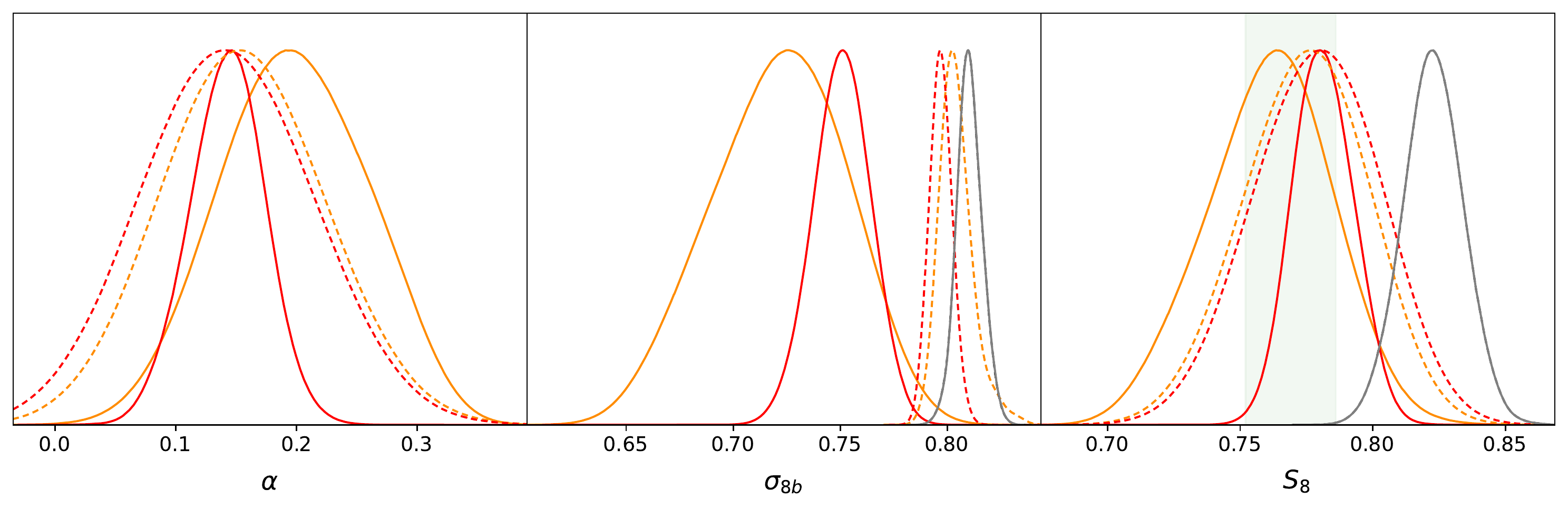}
    \caption{Comparison of 1-D and 2-D distributions of cosmological parameters for both Geodesic and Non-Geodesic with Planck18, JLA and priors (orange curves) and effect of postprocessing the chains with RSD likelihood (red curves). The overlap of solid gray and dashed black curves shows that RSD postprocessing causes no effect on $\Lambda$CDM parameter constraints. The posteriors of $S_8$ parameter shows the agreement of Geodesic results with KiDS-1000 measurements \cite{Heymans_2021} (green shaded area).}
    \label{fig:final}
\end{figure}

\renewcommand{\arraystretch}{1.3}
\begin{table}
    \centering
\begin{tabular}{c|c|c|c|c}
\hline
Parameter               &        Data         & \hspace{.7cm} $\Lambda$CDM  \hspace{.7cm}    & Geodesic CG        & non-Geodesic CG    \\
\hline
 \multirow{2}{*}{ $\alpha$}          & Planck18+JLA      & 0                          & $0.198^{+0.113}_{-0.118}$  & $0.155^{+0.133}_{-0.131}$ \\
                   & Planck18+JLA +RSD & 0                          & $0.143^{+0.061}_{-0.063}$  & $0.143^{+0.117}_{-0.112}$ \\
\hline                   
 \multirow{2}{*}{$\Omega_{m0}$ }    & Planck18+JLA      & $0.309^{+0.012}_{-0.013}$  & $0.226^{+0.045}_{-0.044}$  & $0.242^{+0.049}_{-0.047}$ \\
                   & Planck18+JLA +RSD & $0.309^{+0.012}_{-0.013}$  & $0.247^{+0.025}_{-0.023}$  & $0.247^{+0.055}_{-0.054}$ \\
\hline                   
 \multirow{2}{*}{$\sigma_{8b}$}     & Planck18+JLA      & $0.81\pm0.012$             & $0.722^{+0.058}_{-0.063}$  & $0.804^{+0.018}_{-0.015}$ \\
                   & Planck18+JLA +RSD & $0.81\pm0.012$             & $0.751^{+0.025}_{-0.027}$  & $0.797^{+0.011}_{-0.01}$  \\
\hline                   
\multirow{2}{*}{ $\sigma_{8}$}     & Planck18+JLA      & $0.81\pm0.012$             & $0.879^{+0.039}_{-0.041}$  & $0.864^{+0.043}_{-0.041}$ \\
                   & Planck18+JLA +RSD & $0.81\pm0.012$             & $0.862^{+0.027}_{-0.028}$  & $0.86^{+0.043}_{-0.042}$  \\
\hline                   
\multirow{2}{*}{ $S_{8}$ }          & Planck18+JLA      & $0.823\pm0.023$            & $0.761^{+0.044}_{-0.048}$  & $0.775^{+0.042}_{-0.046}$ \\
                   & Planck18+JLA +RSD & $0.823\pm0.023$            & $0.781\pm0.022$            & $0.779^{+0.042}_{-0.045}$ \\
\hline                   
\multirow{2}{*}{ $H_{0}$ }          & Planck18+JLA      & $67.771^{+0.965}_{-0.899}$ & $70.664^{+1.887}_{-1.932}$ & $70.09^{+1.937}_{-1.92}$  \\
                   & Planck18+JLA +RSD & $67.771^{+0.965}_{-0.899}$ & $69.894^{+1.051}_{-1.146}$ & $69.9^{+1.922}_{-1.948}$  \\
\hline                   
\multirow{2}{*}{ $n_{s}$}           & Planck18+JLA      & $0.968\pm0.008$            & $0.963\pm0.008$            & $0.964\pm0.008$           \\
                   & Planck18+JLA +RSD & $0.968\pm0.008$            & $0.964\pm0.008$            & $0.964\pm0.008$           \\
\hline                   
\multirow{2}{*}{ $\tau_{reio }$ }                  & Planck18+JLA      & $0.057^{+0.016}_{-0.014}$  & $0.056\pm0.015$            & $0.057^{+0.015}_{-0.014}$ \\
                   & Planck18+JLA +RSD & $0.057^{+0.016}_{-0.014}$  & $0.058^{+0.016}_{-0.015}$  & $0.057^{+0.016}_{-0.014}$ \\
\hline                   
\multirow{2}{*}{ $ln10^{10}A_{s }$} & Planck18+JLA      & $3.048^{+0.03}_{-0.027}$   & $3.052^{+0.029}_{-0.03}$   & $3.051^{+0.03}_{-0.027}$  \\
                   & Planck18+JLA +RSD & $3.048^{+0.03}_{-0.027}$   & $3.054^{+0.03}_{-0.029}$   & $3.05^{+0.03}_{-0.027}$   \\
\hline
\end{tabular}
\caption{Mean values and $2\sigma$ credible intervals for cosmological parameters obtained with MontePython chains for Planck18, JLA SNe Ia Compilation and priors. Also, in the second row of each parameter, the results obtained after postprocessing the chains with RSD likelihood.}
    \label{tab:final_table}

\end{table}
\renewcommand{\arraystretch}{1}

\begin{table}
    \centering
\begin{tabular}{l |c | c}
\hline
Data                  &   Geodesic Chaplygin &   non-Geodesic Chaplygin \\
\hline
 RSD              &                -0.05 &                    -0.05 \\
 RSD+JLA          &                -0.46 &                    -0.46 \\
 RSD+Pantheon     &                -0.18 &                    -0.04 \\
 Planck18+JLA     &                -9.84 &                   -10.04 \\
\hline
\end{tabular}
\caption{The values of $\Delta(\chi^2_{min})$ for each dataset, comparing both Chaplygin models against the $\Lambda$CDM , where $\chi^2\equiv -2$ loglikelihood.   }
    \label{tab:my_label}
\end{table}

\subsection{Results}

The constraints obtained with only RSD data are represented as the black curves on figure \ref{fig:standard} for $\Lambda$CDM model and figure \ref{fig:rsd_sne} for both cases of the gCg discussed here. Combined with JLA or Pantheon SNe Ia compilation, tighter parameter constraints are obtained. It is shown on table \ref{tab:results_RSD_SNe} that the inclusion of JLA data does not affect much $\alpha$ mean value in the non-geodesic scenario, but a considerable change is noted when using Pantheon data. On the other hand the opposite happens when the latter and the former are included in the geodesic. Besides RSD data and both SNeIa data used show a preference for a flux of energy from dark energy to dark matter, Pantheon dataset deviates the curves towards the $\Lambda$CDM model. No relevant differences between geodesic and non geodesic models where observed apart from slightly tighter constraints for the latter, what could be addressed to divergences when using certain set of parameters, drawing forbidden regions. The preference towards the standard model when using Pantheon dataset is expected once nuisance parameters are fixed on $\Lambda$CDM best-fit values, and also a lower set of free parameters tends to return tigther constraints on the remanent parameters.

We extend our results by also considering CMB data. In Fig. \ref{fig:full}, with orange curves, we show that Planck+JLA likelihoods favour positive values of $\alpha$ for both the geodesic (left panel) and for the non geodesic (right panel), leading to lower values of $\Omega_{m0}$. 

While our geodesic scenario result is an updated analysis and is compatible with previous results \cite{Benetti:2019lxu, Benetti:2021div}, the non-geodesic case has never been analysed in the literature. In this case we note a sligthly higher value for the $\alpha$ parameter with broader errors.
Also, the Planck18+JLA chains are post-processed with RSD likelihood generating the red curves on Figures \ref{fig:full} and \ref{fig:final}.

Table \ref{tab:final_table} summarizes the mean values and $2\sigma$ credible regions for the parameters of interest. 
The parameters of the $\Lambda$CDM model are insensitive to the  RSD data when combined with Planck18+JLA. This behavior is not observed when we analyse the interacting model. The amplitude of the baryon density fluctuation $\sigma_{8b}$ is slightly larger for the non- geodesic than for the geodesic. As a consequence, the inclusion of RSD yields a shift of $\alpha$ to lower values, leading to slightly higher values of $\Omega_{m0}$. Thus, except for the $\sigma_{8b}$ values, the cosmological parameters resulting from the joint analysis are practically the same for both scenarios. Through the equation ($\ref{sui00}$) it is possible to find that the baryon power spectrum for non-geodesic case is $13\%$ greater than the geodesic one. This difference increases for $28\%$ if only the analysis of Planck18+JLA is considered.  We also note that the $\Lambda$CDM model ($\alpha=0$) is ruled out with and without RSD, with a clear preference for a positive $\alpha$, corresponding to annihilated dark matter.  

It is quite interesting that positive values of $\alpha$ move matter density parameter to lower values, what allows a better agreement for the $S_8$ $(\equiv\sigma_8\sqrt{\Omega_{m0}/0.3})$ parameter, besides the preferred higher $\sigma_8$ values. These behaviour is emphasized on Figure \ref{fig:final}, where the KiDS-1000 measurements of $S_8$ parameter \cite{Heymans_2021} are indicated in the green shaded area.

\section{Final Remarks}

In this work we investigated the evolution of dark matter perturbations in the context of an interacting dark sector corresponding to a decomposed Chaplygin gas model. We have considered two distinct scenarios for the covariant energy-momentum transfer. In the first, the energy transfer follows the dark matter $4$-velocity. In this simplest case the momentum transfer is zero, which implies a homogeneous vacuum energy. We recover the scale-independent second order differential equation for the density contrast, showing that, compared to the $\Lambda$CDM cosmology, the growth rate is suppressed for $\alpha < 0$ due to the homogeneous creation of dark matter, and enhanced when $\alpha > 0$. 

In the second scenario, the momentum transfer is determined by the gradient of the vacuum perturbations, which is proportional to the scalar expansion, $\delta\rho^c_V\propto\delta\Theta^c$. The dynamics in this case is reduced to a single scale-dependent second order equation. We are able to evaluate the size of vacuum energy perturbations compared with the dark matter ones, determining the evolution of the growth rate for diverse gCg background solutions. The vacuum perturbations show to be negligible on scales inside the horizon. 

To explore the bounds of the cosmological parameters of these two scenarios, we use the observational datasets from Planck18, Type Ia Supernova and redshift-space distortions. We used the growth rate of baryons $f_b$ combined with the $\sigma_{8b}$  to account redshift-space distortions inferred from peculiar velocities. 

Furthermore, we have updated the results of \cite{W. Luciano} and \cite{Benetti:2021lxu} using Planck18, JLA and RSD observational datasets. Our results are shown in table 3. One can note that the included RSD data used in our analysis do not change the mean value $H_0$ estimates for the $\Lambda$CDM cosmology. The discrepancy on $H_0$ between the $\Lambda$CDM prediction $H_0=67.771\pm^{0.965}_{0.889}$ with respect to the latest Hubble constant measurement by Hubble Space Telescope and the SH0ES Team \cite{A} is about 3.8 $\sigma$.

 In the presence of a non-zero energy transfer between dark matter and the vacuum the uncertainty on $H_0$ inferred from the combination of Planck+JLA+RSD datasets decreased. For the geodesic scenario the bounds on $H_0$ reduce the tension to 2.0 $\sigma$ while for the non geodesic, where the error is about twice as large, one finds 1.4 $\sigma$. In both scenarios the data prefer a positive values of $\alpha=0.143$ and a lower values of $S_8$ with good agreement with KiDS-1000. 
 
 In conclusion, our analysis shows an interesting agreement between the analysed models and the data, as well as a reduction in the cosmological tensions, that are one of the most heated debates of current cosmology. Finally, we note a clear indication for an energy flux from dark matter to dark energy, being $\alpha$ positive and also different from zero at 3$\sigma$.

\section*{Acknowledgements}

The authors are thankful to Saulo Carneiro and Winfried Zimdahl for a critical reading. 
M.B. acknowledge Istituto Nazionale di Fisica Nucleare (INFN), sezione di Napoli, iniziativa specifica QGSKY.
We also acknowledge the use of Monte Python package. This work was developed thanks to the use of the National Observatory Data Center (CPDON).


\begin{thebibliography}{30}

\bibitem{Astier} P. Astier \emph{et al.}, \emph{The Supernova Legacy Survey: measurement of $\Omega_M$, $\Omega_{\Lambda}$ and $w$ from the first year data set}, \emph{Astronomy \& Astrophysics} \textbf{447}, 31 (2006).

\bibitem{riess} A. G. Riess \emph{et al.}, \emph{Type Ia Supernova Discoveries at $z>1$ From the Hubble Space Telescope: Evidence for Past Deceleration and Constraints on Dark Energy Evolution}, \emph{Astrophys. J.} \textbf{607}, 665 (2004).

\bibitem{perl} S. Perlmutter \emph{et al.}, \emph{Measurements of $\Omega$ and $\Lambda$ from 42 High-Redshift Supernovae}, \emph{Astrophys. J.} \textbf{517}, 565 (1999).

\bibitem{spergel} D. N. Spergel \emph{et al.} (WMAP), \emph{First-Year Wilkinson Microwave Anisotropy Probe (WMAP) Observations: Determination of Cosmological Parameters}, \emph{Astrophys. J. Suppl.} \textbf{148}, 175 (2003).

\bibitem{g} G. Hinshaw \emph{et al.} (WMAP), \emph{First Year Wilkinson Microwave Anisotropy Probe (WMAP) Observations: Angular Power Spectrum}, \emph{Astrophys. J. Suppl.} \textbf{148}, 135 (2003).

\bibitem{ade} Ade P. A. R. \emph{et al.}, \emph{Planck 2015 results. XIII. Cosmological parameters}, \emph{Astronomy \& Astrophysics} \textbf{594}, A13 (2016).

\bibitem{ade1} Ade P. A. R. \emph{et al.}, \emph{Planck 2015 results. XX. Constraints on inflation}, \emph{Astronomy \& Astrophysics} \textbf{594}, A20 (2016).

\bibitem{telubac} T. Delubac \emph{et al.}, \emph{Baryon acoustic oscillations in the Ly$\alpha$ forest of BOSS DR11 quasars}, \emph{Astronomy \& Astrophysics} \textbf{574}, A59 (2015).

\bibitem{ata} M.~Ata 
\textit{et al.}
\emph{The clustering of the SDSS-IV extended Baryon Oscillation Spectroscopic Survey DR14 quasar sample: first measurement of baryon acoustic oscillations between redshift 0.8 and 2.2},
Mon. Not. Roy. Astron. Soc. \textbf{473}, no.4, 4773-4794 (2018)

\bibitem{tegmark} M. Tegmark \emph{et al.}, \emph{Cosmological parameters from SDSS and WMAP}, \emph{Phys. Rev. D} \textbf{69}, 103501 (2004).

\bibitem{pebles} P. J. E. Peebles, B. Ratra, \emph{The cosmological constant and dark energy}, \emph{Rev. Mod. Phys.} \textbf{75}, 559 (2003).

\bibitem{Padmanaban} T. Padmanabhan, \emph{Cosmological constant - the weight of the vacuum}, \emph{Phys. Rept.} \textbf{380}, 235 (2003).

\bibitem{weinberg02} S. Weinberg, \emph{The cosmological constant problem}, \emph{Rev. Mod. Phys.} \textbf{61}, 1 (1989).

\bibitem{kam} A. Y. Kamenshchik, U. Moschella and V. Pasquier, \emph{An alternative to quintessence}, \emph{Phys. Lett. B} {\bf 511}, 265 (2001).

\bibitem{Fabris} J. C. Fabris, S. V. B. Gon\c{c}alves and P. E. de Souza, \emph{Density perturbations in an Universe dominated by the Chaplygin gas}, \emph{Gen. Rel. Grav.} {\bf 34}, 53 (2002).

\bibitem{Bento} M. C. Bento, O. Bertolami and A. A. Sen, \emph{Generalized Chaplygin gas, accelerated expansion, and dark-energy-matter unification}, \emph{Phys. Rev.D} {\bf 66}, 043507 (2002).

\bibitem{Sandvik:2002jz} H. Sandvik, M. Tegmark, M. Zaldarriaga and I. Waga, \emph{The end of unified dark matter?}, \emph{Phys. Rev. D} {\bf 69}, 123524 (2004).
  
\bibitem{bento} M. C. Bento, O. Bertolami and A. A. Sen, \emph{Revival of the unified dark energy–dark matter model?}, \emph{Phys. Rev.D} {\bf 70}, 083519 (2004).

\bibitem{bilic} N. Bilic, G. B. Tupper and R. D. Viollier, \emph{Unification of dark matter and dark energy: The Inhomogeneous Chaplygin gas}, \emph{Phys. Lett. B} {\bf 535}, 17 (2002).

\bibitem{JSA} J. S. Alcaniz, D. Jain and A. Dev, \emph{High - redshift objects and the generalized Chaplygin gas}, \emph{Phys. Rev. D} {\bf 67}, 043514 (2003).

\bibitem{ca} S. Carneiro, M. A. Dantas, C. Pigozzo and J. S. Alcaniz, \emph{A cosmological concordance model with dynamical vacuum term}, \emph{Phys. Rev. D} {\bf 77}, 083504 (2008)
\bibitem{amendola} L. Amendola, I. Waga and F. Finelli, \emph{Observational constraints on silent quartessence}, \emph{JCAP} \textbf{0511}, 009 (2005).

\bibitem{Zi} W. Zimdahl and J. C. Fabris, \emph{Chaplygin gas with non-adiabatic pressure perturbations}, \emph{Class. Quant. Grav.} {\bf 22}, 4311 (2005).

\bibitem{Wa} Y. Wang \emph{et al.}, \emph{Cosmological constraints on a decomposed Chaplygin gas}, \emph{Phys. Rev. D} {\bf 87}, 083503 (2013).

\bibitem{Ba} H. A. Borges \emph{et al.}, \emph{Non-adiabatic Chaplygin gas}, \emph{Phys. Lett.B}  {\bf 727}, 37 (2013).

\bibitem{DW} D. Wands, J. De-Santiago and Y. Wang, \emph{Inhomogeneous vacuum energy}, \emph{Class. Quant. Grav.} {\bf 29}, 145017 (2012).
\bibitem{DW1} Chakkrit Kaeonikhom, Hooshyar Assadullahi, Jascha Schewtschenko and David Wands, \emph{Observational constraints on interacting vacuum energy with linear interactions}, arXiv: 2210.05363.

\bibitem{W} W. Zimdahl \emph{et al.}, \emph{Non-adiabatic perturbations in decaying vacuum cosmology}, \emph{JCAP} \textbf{1104}, 028 (2011).

\bibitem{JB} J. M. Bardeen, \emph{Gauge-invariant cosmological perturbations}, \emph{Phys. Rev. D} {\bf 22}, 1882 (1980).

\bibitem{C} J. S. Alcaniz \emph{et al.}, \emph{A cosmological concordance model with dynamical vacuum term},  \emph{Phys. Lett. B} {\bf 716} 165-170 (2012).

\bibitem{CPS} C. Pigozzo, S. Carneiro, J. S. Alcaniz, H. A. Borges, and J. C. Fabris, \emph{Evidence for cosmological particle creation?}, \emph{JCAP} {\bf 05}, 022 (2016).

\bibitem{W. Luciano} R. F. Vom Marttens \emph{et al.}, \emph{Does a generalized Chaplygin gas correctly describe the cosmological dark sector}, \emph{Phys. Dark Univ.} {\bf15}, 114 (2017). 

\bibitem{HW} Humberto A. Borges and David Wands, \emph{Growth of structure in interacting vacuum cosmologies}, \emph{Phys.Rev.D} {\bf 101}, 103519 (2020).

\bibitem{MH} V. F. Mukhanov, H. A. Feldman, H. R. Brandenberger, \emph{Theory of cosmological perturbations}, \emph{Phys. Rep}. {\bf 215}, 203 (1992).


\bibitem{Benetti:2019lxu} 
M.~Benetti {\it{et al}}.,
\emph{Looking for interactions in the cosmological dark sector},
  \emph{JCAP} {\bf 12}, 023 (2019).


\bibitem{Benetti:2021lxu} 
M.~Benetti {\it{et al}}.,
\emph{Dark sector interactions and the curvature of the Universe in light of Planck's 2018 data},
  \emph{JCAP} {\bf 08}, 014 (2021).
\bibitem{Chan} Chan-Gyung Park, Jai-chan Hwang, Jae-heon Lee, Hyerim Noh, \emph{Roles of Dark Energy Perturbations in Dynamical Dark Energy Models: Can We Ignore Them?}, \emph{Phys. Rev. Lett.} {\bf 103}, 151303 (2009).

\bibitem {rds} S. Carneiro, H. A. Borges, \emph{Dynamical system analysis of interacting models}, \emph{Gen. Rel. Grav.} {\bf 50}, 129 (2018). 

\bibitem{skara_tension_2020} Skara, F. and Perivolaropoulos, L.,\emph{Tension of the $E_G$ statistic and RSD data with Planck/$\Lambda$CDM and implications for weakening gravity}, \emph{Physical Review D} {\bf 6} V 101, (2020).


\bibitem{Betoule:2014frx} 
  M.~Betoule {\it et al.} [SDSS Collaboration],
  \emph{Improved cosmological constraints from a joint analysis of the SDSS-II and SNLS supernova samples},
  \emph{Astronomy \& Astrophysics}  {\bf 568}, A22 (2014).

\bibitem{Scolnic:2017caz} 
  D.~M.~Scolnic {\it et al.},
 \emph{The Complete Light-curve Sample of Spectroscopically Confirmed SNe Ia from Pan-STARRS1 and Cosmological Constraints from the Combined Pantheon Sample},
  \emph{Astrophys. J.}  {\bf 859}, n. 2, 101 (2018).


\bibitem{Taddei:2016iku} 
  L.~Taddei, M.~Martinelli and L.~Amendola,
  \emph{Model-independent constraints on modified gravity from current data and from the Euclid and SKA future surveys},
  \emph{JCAP} {\bf 1612}, 032 (2016).

  
  

\bibitem{Aghanim:2019ame} 
  N.~Aghanim {\it et al.} [Planck Collaboration],
  \emph{Planck 2018 results. V. CMB power spectra and likelihoods},
  \emph{Astronomy \& Astrophysics}  {\bf 641}, A5 (2020).



\bibitem{feroz_multinest_2009}  Feroz, F. \emph{et al.}, \emph{MULTINEST: an efficient and robust Bayesian inference tool for cosmology and particle physics}, \emph{Monthly Notices of the Royal Astronomical Society}, {\bf 398},  1601 (2009).

\bibitem{Buchner} Buchner, J. \emph{et al.}, \emph{X-ray spectral modelling of the AGN obscuring region in the CDFS: Bayesian model selection and catalogue}, \emph{Astronomy \& Astrophysics} {\bf 564}, A125 (2014).


\bibitem{priorH0_Riess}  Riess, Adam G. \emph{et al.}, \emph{Large Magellanic Cloud Cepheid Standards Provide a $1\%$ Foundation for the Determination of the Hubble Constant and Stronger Evidence for Physics beyond $\Lambda$CDM},  \emph{ApJ} {\bf 876}, 85 (2019).

\bibitem{prioOmgeabh2_Cooke} Cooke, Ryan J. \emph{et al.}, \emph{One Percent Determination of the Primordial Deuterium Abundance}, \emph{ApJ} {\bf  855}, 102 (2018).

\bibitem{Blas:2011rf} 
  D.~Blas, J.~Lesgourgues and T.~Tram,
  \emph{The Cosmic Linear Anisotropy Solving System (CLASS) II: Approximation schemes},
  \emph{JCAP} {\bf 1107}, 034 (2011).
  
\bibitem{Audren:2012wb} 
  B.~Audren, J.~Lesgourgues, K.~Benabed and S.~Prunet,
  \emph{Conservative Constraints on Early Cosmology: an illustration of the Monte Python cosmological parameter inference code},
 \emph{JCAP} {\bf 1302}, 001 (2013).

  
  \bibitem{Aghanim:2015xee} 
  N.~Aghanim {\it et al.} [Planck Collaboration], \emph{Planck 2015 results. XI. CMB power spectra, likelihoods, and robustness of parameters},
 \emph{Astronomy \& Astrophysics}  {\bf 594}, A11 (2016).

 
 
\bibitem{Heymans_2021} Heymans, C. {\it{et al.}}, \emph{KiDS-1000 Cosmology: Multi-probe weak gravitational lensing and spectroscopic galaxy clustering constraints}, \emph{Astronomy \& Astrophysics} {\bf 646}, A140 (2021).
 

\bibitem{Benetti:2021div}
M.~Benetti {\it{et al}}.,
\emph{Dark sector interactions and the curvature of the Universe in light of Planck's 2018 data}, \emph{JCAP} {\bf 2108}, 014 (2021).


\bibitem{A} A.G. Riess et al., \emph{A Comprehensive Measurement of the Local Value of the Hubble Constant with 1 km $s^{-1}$ $Mpc^{-1}$ Uncertainty from the Hubble Space Telescope and the SH0ES Team}, \emph{Astrophys. J. Lett}. {\bf 934} (2022).



\end{thebibliography}
\end{document}